\documentclass[aps,prd,twocolumn,amsmath,amssymb,floatfix]{revtex4-2}

\usepackage{graphicx}
\usepackage{bm}
\usepackage{hyperref}
\usepackage{lipsum}
\usepackage{siunitx}
\usepackage{subfig}
\usepackage{comment}

\begin{document}

\preprint{APS/123-QED}

\title{The all-charm tetraquark and its contribution \\ to two-photon processes}

\author{Panagiotis Kalamidas}
\author{Marc Vanderhaeghen}%
\affiliation{Institut f\"ur Kernphysik and PRISMA\textsuperscript{+} Cluster of Excellence,
Johannes Gutenberg-Universit\"at, Mainz, D-55099, Germany}

\date{\today}

\begin{abstract}
    Prompted by several potential tetraquark states that have been reported by the LHCb, CMS, and ATLAS Collaborations in the di-$J/\Psi$ and $J/\Psi \Psi(2S)$ spectra, we investigate their contribution to two-photon processes.
    Within a non-relativistic potential model for the all-charm tetraquark states, we calculate the two-photon decay widths of these states and test model-independent sum rule predictions. 
    Imposing such sum-rule constraints allows us to predict the light-by-light scattering cross sections in a consistent way and check if any excess, in comparison to the Standard Model prediction, as reported by ongoing ATLAS experiments, can be attributed to intermediate exotic states.
\end{abstract}

\maketitle

\section{\label{sec:Introduction}Introduction}

In recent years, several experiments have revealed a plethora of potential tetraquark states. They can be composed of only light quarks, heavy quarks, or a mixture of both.
A candidate of the second category is the all-charm tetraquark, which is a $cc\overline{c}\overline{c}$ bound state.
The first indication of an all-charm tetraquark state was discovered by the LHCb Collaboration as a resonance around $\SI{6.9}{\giga\eV}$, $X(6900)$ in the di-$J/\Psi$ invariant mass spectrum \cite{LHCb_2020}.
Shortly after, the ATLAS Collaboration confirmed the excess observed in the LHCb analysis and described the data using a three resonance model \cite{ATLAS_2023}.
Simultaneously, the CMS Collaboration also confirmed the excess \cite{CMS_2024} and used a three resonance description of the observed structures through three Breit-Wigner resonances and their interferences.
The resulting resonances and their corresponding widths are shown in Table~\ref{tab:CMS_masses}.

\begin{table}[h!]
    \caption{Breit-Wigner resonance masses and widths of $cc\overline{c}\overline{c}$ states seen in di-$J/\Psi$ spectrum with interference from Ref.~\cite{CMS_2024}}
    \begin{ruledtabular}
    \begin{tabular}{ccc}
        resonance $i$ & $M_{i} (\SI{}{\giga\eV})$ & $\Gamma_{i} (\SI{}{\giga\eV})$ \\ \colrule
         & & \\[-1.0em]
        $1$ & $6.638 ^{+0.043+0.016}_{-0.038-0.031}$ & $0.440 ^{+0.230+0.110}_{-0.200-0.240}$ \\
         & & \\[-1.0em]
        $2$ & $6.847 ^{+0.044+0.048}_{-0.028-0.020}$ & $0.191 ^{+0.066+0.025}_{-0.049-0.017}$ \\
         & & \\[-1.0em]
        $3$ & $7.134 ^{+0.048+0.041}_{-0.025-0.015}$ & $0.097 ^{+0.040+0.029}_{-0.029-0.026}$ \\
    \end{tabular}
    \end{ruledtabular}
    \label{tab:CMS_masses}
\end{table}

In parallel to this, in 2017 the ATLAS Collaboration found the first evidence of Light-by-Light (LbL) scattering \cite{ATLAS_2017} in heavy-ion collisions.
Subsequently, in a more recent analysis \cite{ATLAS_2021} of all data from Run 2 at the LHC, this discovery was strengthened to reach a significance of $8.2 \ \sigma$.
The resulting data in the diphoton electroproduction cross section have indicated a potential discrepancy with the predicted value of the Standard Model, especially in the $5$ to $\SI{10}{\giga\eV}$ region.

On the theoretical side, exotic multiquark states have been discussed for decades.
The first work on an all-charm tetraquark configuration was made in Refs.~\cite{Iwasaki_1975, Iwasaki_1976, Iwasaki_1977}.
Since then, many approaches have been used to predict the mass spectra of the resulting states.
One of the main ways to tackle this problem is by using a diquark picture for the quark interactions inside the bound state.
In combination with a potential, such as the Cornell potential \cite{Cornell_1975}, it can be used to predict the bound states and their characteristics.

LbL scattering, on the other hand, is one of the earliest processes predicted by QED \cite{Euler_1936,*Heisenberg_1936}.
Model-independent sum rules for LbL processes \cite{Pascalutsa:2010sj,Pascalutsa:2012pr} are particularly useful in this context.
Furthermore, work performed in \cite{Biloshytskyi_2022} bridges all-charm tetraquark searches with LbL studies. 
In particular, it suggests that an all-charm tetraquark candidate, in this case the $X(6900)$ could be responsible for the excess of LbL events seen by the ATLAS Collaboration.

In this work, we follow and expand on the description of the diquark potential model of~\cite{Debastiani_2019} to calculate the predicted all-charm tetraquark spectrum, including the $D$ states as well.
For the resulting states, we then calculate the two-photon decay widths for the first time.
Our aim is to check whether these states allow us to explain the perceived experimental excess in the LbL data.
Putting these results against the model-independent sum rule for real photons, we propose a modified potential model for these states that can potentially accommodate the existing experimental data.

In Section~\ref{sec:Model} we describe the diquark potential model used in this work and calculate the resulting spectrum.
After that, in Section~\ref{sec:Two-photon} we first calculate the perturbative two-photon widths for axial-vector diquarks and use them to predict the widths of the all-charm tetraquark bound states.
In Section~\ref{sec:Results and discussion}, we introduce the modified potential model which is fitted to the experimental all-charm tetraquark states and is then compared with other works. 
Finally, Section~\ref{sec:Conclusion} consists of the main conclusions and an outlook.

\section{\label{sec:Model}Tetraquark Potential Model}

\begin{figure}[h!]
    \includegraphics[width=\linewidth]{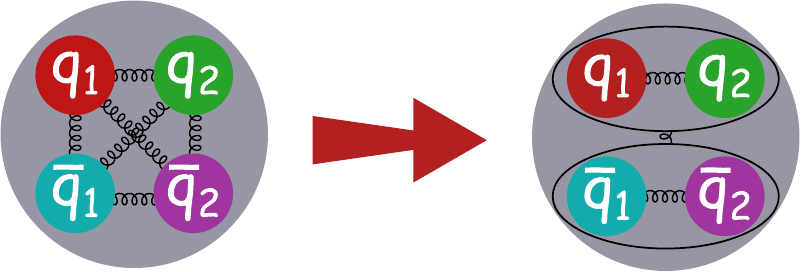}
    \caption{Schematic representation of the diquark picture considering one-gluon interactions. In this framework, the four-body problem (left) is converted to three simpler two-body problems with one-gluon interactions (right). The quarks $q_1$, $q_2$ form the diquark and $\overline{q}_1$, $\overline{q}_2$ form the antidiquark.}
    \label{fig:scem_diquark}
\end{figure}

Considering all interactions between four valence quarks in a tetraquark state requires solving a complicated four-body problem.
A common approximation is therefore to introduce a diquark picture.
In such a framework, two quarks are bound to form a diquark quasiparticle, and antiquarks form the antidiquark, respectively.
Two of those composite particles are treated as compact, and an attractive interaction between them gives rise to the bound tetraquark state.
Consequently, in such a picture, the complicated four-body problem is reduced to three two-body ones.
A pictorial representation of this concept is shown in Figure~\ref{fig:scem_diquark}.

The color content of an all-charm tetraquark configuration follows from the $SU(3)_c$ representations.
Two quarks, described by a color triplet representation, are combined as
\begin{equation}
    \mathbf{3} \otimes \mathbf{3} = \mathbf{6} \oplus \mathbf{\overline{3}}.\label{eq:3_3}
\end{equation}
The sextet configuration is repulsive, while the anti-triplet is attractive.
For this analysis, only the attractive anti-triplet will be considered.
Similarly, antiquarks couple as,
\begin{equation}
    \mathbf{\overline{3}} \otimes \mathbf{\overline{3}} = \mathbf{\overline{6}} \oplus \mathbf{3},\label{eq:3bar_3bar}
\end{equation}
likewise only the triplet will be used.
The diquark anti-triplet and the antidiquark triplet are then combined as,
\begin{equation}
    \mathbf{3} \otimes \mathbf{\overline{3}} = \mathbf{8} \oplus \mathbf{1}.\label{eq:3_3bar}
\end{equation}
We will only consider resulting color singlet tetraquark bound states, in Eq.~\eqref{eq:3_3bar}, which is the most attractive configuration under one-gluon exchange.

Due to the relatively large mass of the charm quark, a non-relativistic framework will be adopted.
Consequently, for each of the three two-body problems, we will solve the reduced Schr\"odinger equation, which is
\begin{equation}
        \left[ \frac{1}{2m} \left( - \frac{\text{d}^2}{\text{d} r^2} + \frac{l (l+1)}{r^2} \right) + \mathcal{V}(r) \right] u(r) = E u(r),\label{eq:schroedinger}
\end{equation}
where $u(r)$ corresponds to the reduced wavefunction,
\begin{equation}
    u(r) = \frac{R(r)}{r}\label{eq:u(r)}.
\end{equation}
The interaction between the charm quarks and between the charmed diquarks is then approximated by a potential of Cornell type \cite{Cornell}, consisting of a central potential, complemented by spin-spin, spin-orbit and tensor terms.
The considered potential given by \cite{Debastiani_2019},
\begin{equation}
    \begin{aligned}
        V =& \mathcal{V}_C (r) + \mathcal{V}_{SS} (r) \delta^3 (\mathbf{r}) \mathbf{S}_1 \cdot \mathbf{S}_2 + \mathcal{V}_{LS} (r) \mathbf{L \cdot S} \\
        &+ \mathcal{V}_{T} (r) \biggl( \frac{(\mathbf{S}_1 \cdot \mathbf{r})(\mathbf{S}_2 \cdot \mathbf{r})}{r^2} - \frac{1}{3} \mathbf{S}_1 \cdot \mathbf{S}_2 \biggr),
    \end{aligned}
    \label{eq:generalv}
\end{equation}
with
\begin{equation}
    \begin{aligned}
    \mathcal{V}_C (r) =& \kappa_s \frac{\alpha_s}{r} + b r,\\
    \mathcal{V}_{SS} (r) =& - \frac{8 \kappa_s \alpha_s \pi}{3 m^2},\\
    \mathcal{V}_{LS} (r) =& - \frac{3 \kappa_s \alpha_s}{2m^2} \frac{1}{r^3} - \frac{b}{2m^2} \frac{1}{r},\\
    \mathcal{V}_{T} (r) =& - \frac{12 \kappa_s \alpha_s}{4 m^2} \frac{1}{r^3}, \\
    \label{eq:potential_specific}
\end{aligned}
\end{equation}
where $r$ corresponds to the relative distance between the interacting particles.
Eq.~\eqref{eq:potential_specific} contains four parameters, $\kappa_s$ is the color factor resulting from the one-gluon exchange potential, $\alpha_s$ is the strong coupling constant, $b$ is the slope parameter from the confinement potential, and $m$ is the mass of the constituents (quarks or diquarks and their antiparticles).

The Cornell potential requires a numerical solution for the reduced Schr\"odinger equation.
To avoid the singular $\delta$ function in the spin-spin interaction potential, a Gaussian smearing function is commonly used instead.
Introducing a smearing parameter $\sigma$, the spin-spin interaction becomes,
\begin{equation}
    V_{SS} = \mathcal{V}_{SS} (r) \left( \frac{\sigma}{\sqrt{\pi}} \right)^3 e^{- \sigma^2 r^2} \mathbf{S_1 \cdot S_2}.\label{eq:vss}
\end{equation}

The color factor $\kappa_s$ depends on the color configuration and can be calculated from Feynman tree-level graphs.
In Table~\ref{tab:color_factor} we have listed the factor $\kappa_s$ for a quark-quark antitriplet configuration (the same for the antiquark-antiquark triplet) and a diquark-antidiquark singlet configuration.

\begin{table}[h!]
    \caption{\label{tab:color_factor} Table of color factors $\kappa_s$ for the different configurations considered.}
    \begin{tabular}{cc}
    \hline\hline
        \textrm{configuration} & $\kappa_s$ \\
        \colrule
        \textrm{singlet} & $-4/3$ \\
        \textrm{anti-triplet} & $-2/3$ \\
        \hline\hline
    \end{tabular}
\end{table}

As the color triplet state of a diquark composed of two charm quarks is anti-symmetric, its spatial-spin wavefunction has to be symmetric to comply with the Pauli principle.
This leads for ground-state diquarks (in an s-wave orbital state) to a symmetric spin-1 configuration.
As a result, the ground state $cc$ or $\overline{c}\overline{c}$ states are spin-$1$ axial-vector diquark states.
Subsequently, the diquark and antidiquark spins combine to give three different tetraquark configurations, corresponding to the total spin $S=0,1,2$.
The total spin $S$ is combined with the relative angular momentum between the diquark and the anti-diquark to give the total angular momentum $J$ of the tetraquark.
When the angular momenta operators act upon wavefunctions of tetraquark states with definite quantum numbers, they generate the following expectation values for the operators of $V_{SS}$ and $V_{LS}$,
\begin{align}
    \langle \mathbf{S}_1 \cdot \mathbf{S}_2 \rangle \nonumber = \frac{1}{2} \{ S(S+1))-S_1 (S_1 +1)-S_2 (S_2 +1)\},
\end{align}
and
\begin{align}
    \langle \mathbf{S \cdot L} \rangle \nonumber = \frac{1}{2} \{J(J+1)-L(L+1)-S(S+1)\}.
\end{align}

On the other hand, the tensor operator of $V_{T}$ is more involved.
Our approach and notation are based on Ref.~\cite{Debastiani_2019}.
The tensor operator for the tetraquark is defined as
\begin{equation}
    \mathbf{T}_{d\overline{d}}= 12 \left( \frac{(\mathbf{S}_d \cdot \mathbf{r})(\mathbf{S}_{\overline{d}} \cdot \mathbf{r})}{r^2} - \frac{1}{3} \mathbf{S}_d \cdot \mathbf{S}_{\overline{d}} \right)\label{eq:tensor_formula}.
\end{equation}
Eq.~\eqref{eq:tensor_formula} can be expressed as rank-2 tensors that independently act on each diquark as follows:
\begin{equation}
    \mathbf{T}_{d\overline{d}} = 4 \left( T_0 + T'_0 +T_1 +T_{-1} + T_2 + T_{-2} \right),\label{eq:diquark_tensor}
\end{equation}
with
\begin{equation}
    \begin{aligned}
    T_0    &= 2 \sqrt{\frac{4 \pi}{5}} Y^0_2 (\theta,\phi) S_{dz} S_{\overline{d}z}\\
    T'_0   &= -\frac{1}{4} 2 \sqrt{\frac{4 \pi}{5}} Y^0_2 (\theta,\phi) (S_{d+} S_{\overline{d}-}+S_{d-} S_{\overline{d}+}),\\
    T_1    &= \frac{3}{2} \sqrt{\frac{8 \pi}{15}} Y^{-1}_2 (\theta,\phi) (S_{dz} S_{\overline{d}+}+S_{d+} S_{\overline{d}z}),\\
    T_{-1} &= - \frac{3}{2} \sqrt{\frac{8 \pi}{15}} Y^{1}_2 (\theta,\phi) (S_{dz} S_{\overline{d}-}+S_{d-} S_{\overline{d}z}),\\
    T_2    &= 3 \sqrt{\frac{2 \pi}{15}} Y^{-2}_2 (\theta,\phi) S_{d+} S_{\overline{d}+},\\
    T_{-2} &= 3 \sqrt{\frac{2 \pi}{15}} Y^{2}_2 (\theta,\phi) S_{d-} S_{\overline{d}-}, \label{eq:tensor_specifics}
\end{aligned}
\end{equation}
where $S_{d(\overline{d})z}, \ S_{d(\overline{d})z+}$ and $S_{d(\overline{d})z-}$ correspond to the spin operators for the diquark (antidiquark), $Y^{M_L}_L (\theta, \phi)$ refers to the spherical harmonics with $L$ and $M_L$ being the orbital angular momenta of the tetraquark and its projection.
The only remaining point is to decompose the tetraquark states into diquark spin eigenstates.

The states with $L = 0$ force the tensor factor to be trivially zero.
For $L = 1$, the factors were calculated in~\cite{Debastiani_2019}.
We extend their calculation to $L = 2$, for which the tensor factors are presented in Table~\ref{tab:tensor_factor}.
An example case of this calculation is provided in the Appendix~\ref{ap:tensor_factor}.

\begin{table}[h!]
        \caption{\label{tab:tensor_factor} Matrix elements $\langle T_{d\overline{d}} \rangle$ of the tensor operator for different tetraquark states states with $L = 2$.}
        \begin{tabular}{cc}
        \hline\hline
            $(L,S,J)$ & $\langle T_{d\overline{d}} \rangle$ \\ \hline
            $(2,0,2)$ & $0$       \\
            $(2,1,1)$ & $-104/35$ \\
            $(2,1,2)$ & $12/7$    \\
            $(2,1,3)$ & $-8/7$    \\
            $(2,2,0)$ & $-28/5$   \\
            $(2,2,1)$ & $-68/35$  \\
            $(2,2,2)$ & $132/49$  \\
            $(2,2,3)$ & $32/7$    \\
            $(2,2,4)$ & $-16/7$  \\ \hline\hline
        \end{tabular}
\end{table}

The next step is to solve the Schr\"odinger equation numerically for the potential of Eq.~\eqref{eq:generalv}.
Due to their small relative contribution compared to the central part of the potential, we can treat $V_{LS}$ and $V_{T}$ as perturbations.
As a first approach of producing the all-charm tetraquark spectrum the same parameters as in Ref~\cite{Debastiani_2019} were used:
\begin{equation}
    \begin{gathered}
        \alpha_s = 0.5202,\\
        m_d = \SI{3.1334}{\giga\eV},\\
        b=\SI{0.1463}{\giga\eV\tothe{2}},\\
        \sigma=\SI{1.0831}{\giga\eV}.
        \label{eq:paper_parameters}
    \end{gathered}
\end{equation}

\begin{table*}
    \caption{All-charm tetraquark bound state potential contributions with $1 \, ^3 S_1$ diquarks for states with $L = 2$. The same parameters and the same number of digits were used as in \cite{Debastiani_2019}.}
    \begin{ruledtabular}
        \begin{tabular}{cccccccc}
            $n \, ^{2S+1} {L}_{J}$ & $E_{d \overline{d}} \ ( \SI{}{\giga\eV})$ & $\langle T \rangle\ (\SI{}{\giga\eV})$ & $\langle V_V \rangle\ (\SI{}{\giga\eV})$ & $\langle V_b \rangle\ (\SI{}{\giga\eV})$ & $\langle V_{SS} \rangle\ (\SI{}{\giga\eV})$ & $\langle V_{LS} \rangle\ (\SI{}{\giga\eV})$ & $\langle V_{T} \rangle\ (\SI{}{\giga\eV})$  \\ \hline
            $1\, ^1 D_2$  & $0.5809$  & $0.3591$ & $-0.2424$ & $0.4662$ & $-0.0020$ & $0      $ & $0      $ \\
            $1\, ^3 D_1$  & $0.5819$  & $0.3568$ & $-0.2415$ & $0.4675$ & $-0.0009$ & $-0.0126$ & $-0.0034$ \\
            $1\, ^3 D_2$  & $0.5819$  & $0.3568$ & $-0.2415$ & $0.4675$ & $-0.0009$ & $-0.0042$ & $ 0.0019$ \\
            $1\, ^3 D_3$  & $0.5819$  & $0.3568$ & $-0.2415$ & $0.4675$ & $-0.0009$ & $0.0084 $ & $-0.0013$ \\
            $1\, ^5 D_0$  & $0.5837$  & $0.3526$ & $-0.2398$ & $0.4700$ & $ 0.0009$ & $-0.0239$ & $-0.0061$ \\
            $1\, ^5 D_1$  & $0.5837$  & $0.3526$ & $-0.2398$ & $0.4700$ & $ 0.0009$ & $-0.0200$ & $-0.0021$ \\
            $1\, ^5 D_2$  & $0.5837$  & $0.3526$ & $-0.2398$ & $0.4700$ & $ 0.0009$ & $-0.0120$ & $ 0.0029$ \\
            $1\, ^5 D_3$  & $0.5837$  & $0.3526$ & $-0.2398$ & $0.4700$ & $ 0.0009$ & $0      $ & $ 0.0050$ \\
            $1\, ^5 D_4$  & $0.5837$  & $0.3526$ & $-0.2398$ & $0.4700$ & $ 0.0009$ & $0.0160 $ & $-0.0025$ \\
        \end{tabular}
    \end{ruledtabular}
    \label{tab:pot_contributions}
\end{table*}

The method used to solve Eq.~\eqref{eq:schroedinger} was spatial democratization with the Arnoldi iteration \cite{Arnoldi_1951}.
The solution range was $\SI{0}{\giga\eV\tothe{-1}} \leq r \leq \SI{40}{\giga\eV\tothe{-1}}$ covered with a step of $\SI{0.001}{\giga\eV\tothe{-1}}$ and the radial wave function $R(r)$ was normalized to unity.
The masses of the all-charm tetraquark states were computed by,
\begin{equation}
    \begin{aligned}
        M_T ({}^{2S + 1} {L}_{J}) =& 2 m_d + E_{d \overline{d}} \\
        &  +\langle   V_{LS}(r) +V_{T}(r) \rangle_{d \overline{d}},
    \end{aligned}
\end{equation}
where $E_{d \overline{d}}$ is the energy eigenvalue of the diquark-antidiquark system.
The subscript $d \overline{d}$ states that the calculation of the expectation value uses the diquark eigenfunctions.

Since the problem has already been solved for the $S$- and $P$-states in Ref.~\cite{Debastiani_2019} which we reproduced as a check, we are focusing here on the $D$-states of the spectrum.
The potential contributions for $L = 2$ are presented in Table~\ref{tab:pot_contributions} for the first energy shell.
As for notation, $V_{b}$ corresponds to the linear confinement term, $T$ is the kinetic energy of the bound states and $V_{V}$ is the Coulomb term of Eq.~\eqref{eq:potential_specific}.
The contributions for $V_{LS}$ and $V_{T}$ are significantly smaller than $E_{d \overline{d}}$, thus justifying the assumption made above.
The masses of the all-charm tetraquark bound $D$-states are presented in Table~\ref{tab:masses_paper} for the first four energy levels.
In each shell there are multiple states in a relatively small energy region making the experimental distinction between them potentially difficult.
The numerical solutions also yield the interpolated radial wavefunction $R(r)$, which together with the mass of the tetraquark states enter the further calculations of two-photon production.

\begin{table}[h!]
    \caption{All-charm tetraquark masses with $1^3 S_1$ diquarks for states with $L = 2$. The same parameters and the same number of digits were used as in Ref~\cite{Debastiani_2019}.}
    \begin{ruledtabular}
    \begin{tabular}{ccccc}
        & \multicolumn{4}{c}{$M_T \ (\SI{}{\giga\eV})$}\\ \colrule
        & & & & \\[-1.0em]
       ${}^{2S+1} {L}_{J}$ & $n=1$ & $n=2$ & $n=3$ & $n=4$ \\ \colrule
       ${}^1 D_2$ & $6.8477$ & $7.1460$ & $7.4040$ & $7.6365$ \\
       ${}^3 D_1$ & $6.8328$ & $7.1318$ & $7.3900$ & $7.6227$ \\
       ${}^3 D_2$ & $6.8464$ & $7.1450$ & $7.4030$ & $7.6356$ \\
       ${}^3 D_3$ & $6.8558$ & $7.1545$ & $7.4127$ & $7.6453$ \\
       ${}^5 D_0$ & $6.8204$ & $7.1208$ & $7.3795$ & $7.6125$ \\
       ${}^5 D_1$ & $6.8284$ & $7.1284$ & $7.3869$ & $7.6197$ \\
       ${}^5 D_2$ & $6.8415$ & $7.1409$ & $7.3992$ & $7.6318$ \\
       ${}^5 D_3$ & $6.8555$ & $7.1545$ & $7.4127$ & $7.6453$ \\
       ${}^5 D_4$ & $6.8640$ & $7.1633$ & $7.4218$ & $7.6547$ \\ 
    \end{tabular}
    \end{ruledtabular}
    \label{tab:masses_paper}
\end{table}

\section{\label{sec:Two-photon}Two-photon production of tetraquarks}

\begin{figure}
    \subfloat[]{\includegraphics[width=0.49\linewidth]{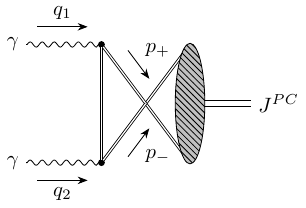}}
    \subfloat[]{\includegraphics[width=0.49\linewidth]{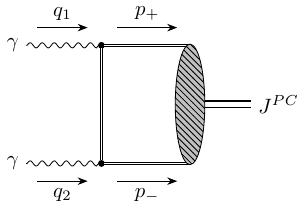}}\\
    \subfloat[]{\raisebox{0mm}{\includegraphics[width=0.49\linewidth]{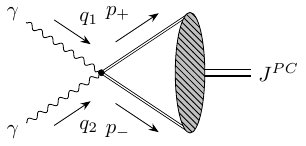}}}
    \caption{Feynman graphs for the two photon to two diquark process with the production of $J^{PC}$ tetraquark state.}
    \label{fig:two_photon_graphs}
\end{figure}

Using the tetraquark wave functions as bound states of a diquark-antidiquark, we are now in a position to calculate the production rate for a tetraquark bound state in the fusion of two photons, or equivalently, the decay width of the state to two photons. 
The resulting Feynman diagrams for computing this process are presented in Figure~\ref{fig:two_photon_graphs}.
Two photons are annihilated to produce diquarks (double line), which form the all-charm tetraquark state with quantum numbers $J^{PC}$.
The matrix element for the two-photon to diquark-antidiquark transition can be calculated with perturbative methods.
The blob indicates the tetraquark wave function, obtained as the diquark-antidiquark bound state.

To describe the vertex for spin-$1$ diquarks interacting with photons a natural choice is the Standard Model $W$-boson coupling, which yields cross sections that respect tree-level unitarity.
The specific Feynman rules can be found in Appendix~\ref{ap:two_photon}.
We work in the center-of-mass frame with the kinematics and the momenta assignments presented in Figure~\ref{fig:kinematics}, with $p_+, \ p_-$ corresponding to the (anti)diquark four-momenta, $q_1, \ q_2$ to the photon four-momenta and where $\theta$ is the angle between the directions of $\mathbf{q}_1$ and $\mathbf{p}_+$: $\cos \theta = \hat{q}_1 \cdot \hat{p}_+$.
Furthermore, $\lambda_+$, $\lambda_-$ correspond to the (anti)diquark helicities and $\lambda_1$, $\lambda_2$ to the photon helicities.

In order to simplify the process, we use the helicity-amplitude method.
We refer for the details of the calculation to the Appendix~\ref{ap:two_photon}, which yields the helicity amplitudes
\begin{widetext}
\begin{equation}
    h_{\lambda_+ \lambda_-}^{\lambda_1 \lambda_2}=\varepsilon_\nu (q_2,\lambda_2) \varepsilon_\mu (q_1,\lambda_1) \varepsilon_\alpha^* (p_+,\lambda_+) \varepsilon_\beta^* (p_-,\lambda_-) \left[ \mathcal{M}^{\alpha \beta, \mu \nu}_a + \mathcal{M}^{\alpha \beta, \mu \nu}_b + \mathcal{M}^{\alpha \beta, \mu \nu}_c \right],\label{eq:amplitude}
\end{equation}
where
\begin{equation}
    \begin{aligned}
        \mathcal{M}^{\alpha \beta, \mu \nu}_a &= \frac{2 i e_d^2}{q_1 \cdot p_-} \biggl[ g^{\alpha \beta} p_-^\mu p_+^\nu + g^{\mu \nu} q_1^\beta q_2^\alpha +g^{\alpha \nu} (- q_2^\mu q_1^\beta -p_-^\mu q_2^\beta) +g^{\beta \mu} (-q_1^\nu q_2^\alpha -p_+^\nu q_1^\alpha)\\
        & \qquad +g^{\alpha \mu} p_+^\nu q_1^\beta + g^{\beta \nu} p_-^\mu q_2^\alpha + g^{\alpha \nu} g^{\beta \mu} \biggl(-\frac{1}{2} q_1 \cdot p_- + q_1 \cdot q_2\biggr)  \biggr],\\
        \mathcal{M}^{\alpha \beta, \mu \nu}_b &= \frac{2 i e_d^2}{q_1 \cdot p_-} \biggl[ g^{\alpha \beta} p_+^\mu p_-^\nu + g^{\mu \nu} q_2^\beta q_1^\alpha +g^{\alpha \mu} (- q_2^\beta q_1^\nu -p_-^\nu q_1^\beta) +g^{\beta \nu} (-q_2^\mu q_1^\alpha -p_+^\mu q_2^\alpha)\\
        & \qquad +g^{\alpha \nu} p_+^\mu q_2^\beta + g^{\beta \mu} p_-^\nu q_1^\alpha + g^{\alpha \mu} g^{\beta \nu} \biggl(-\frac{1}{2} q_1 \cdot p_+ + q_1 \cdot q_2 \biggr)  \biggr],\\
        \mathcal{M}^{\alpha \beta, \mu \nu}_c &= i e_d^2 \biggl[ g^{\alpha \nu} g^{\beta \mu} + g^{\alpha \mu} g^{\beta \nu} -2 g^{\alpha \beta} g^{\mu \nu} \biggr],\label{eq:amplitude_abc}
    \end{aligned}
\end{equation}
\end{widetext}
with $e_d$ being the electric charge of the diquark.

\begin{table}[h!]
    \caption{Helicity amplitude factors for the $\gamma \gamma \to V V$ process appearing in Eq.~\eqref{eq:amplitude_helicity}.}
    \begin{ruledtabular}
    \begin{tabular}{cccc}
          $\lambda_+$&$\lambda_-$  &  $f^{\Lambda=0}_{\lambda_+ \lambda_-} (\beta, \theta)$ & $f^{\Lambda=2}_{\lambda_+ \lambda_-} (\beta, \theta)$ \\ \colrule
         $ 1$  & $ 1$  & $(1+\beta)^2$ & $(1-\beta^2) \sin ^2 \theta $ \\
         $ 1$  & $ -1$ & $0$           & $(1+\cos \theta)^2$ \\
         $ 1$  & $ 0$  & $0$           & $\frac{2 \sqrt{2} m_d}{\sqrt{s}} (1+\cos \theta) \sin \theta$ \\
         $ -1$ & $ 1$  & $0$           & $(1-\cos \theta)^2$ \\
         $ -1$ & $ -1$ & $(1-\beta)^2$ & $(1-\beta^2) \sin ^2 \theta $ \\
         $ -1$ & $ 0$  & $0$           & $\frac{2 \sqrt{2} m_d}{\sqrt{s}} (1-\cos \theta) \sin \theta$ \\
         $ 0$  & $ 1$  & $0$           & $-\frac{2 \sqrt{2} m_d}{\sqrt{s}} (1-\cos \theta) \sin \theta$ \\
         $ 0$  & $ -1$ & $0$           & $-\frac{2 \sqrt{2} m_d}{\sqrt{s}} (1+\cos \theta) \sin \theta$ \\
         $ 0$  & $ 0$  & $(1-\beta^2)$ & $-(2-\beta^2) \sin ^2 \theta $
    \end{tabular}
    \end{ruledtabular}
    \label{tab:helicity_amplitudes}
\end{table}

Using invariance under parity, only two out of the four helicity combinations for the photons are independent.
Those amplitudes are $h_{\lambda_+ ,\lambda_- }^{+1,\pm 1} (\beta, \theta)$, corresponding to the helicity $\Lambda \equiv \lambda_1 - \lambda_2$ of the two-photon state, 
which is either $0$ or $2$.
The amplitude for different $\lambda_+$ and $\lambda_-$ pairs can be reduced to the following form,
\begin{equation}
    h_{\lambda_+ \lambda_-}^{+1 \pm 1} (\beta, \theta)= \frac{2ie_d^2}{1-\beta^2 \cos^2 \theta}  f^{\Lambda=0,2}_{\lambda_+ \lambda_-} (\beta, \theta),\label{eq:amplitude_helicity}
\end{equation}
where $f^{\Lambda=0,2}_{\lambda_+ \lambda_-} (\beta, \theta)$ are helicity amplitude factors dependening on the angle $\theta$ and the diquark velocity in the c.m. frame  
\begin{equation}
    \beta = \sqrt{1-\frac{4 m_d^2}{s}},
\end{equation}
with $s = (q_1 + q_2)^2$. 
The factors $f^{\Lambda=0,2}_{\lambda_+ \lambda_-} (\beta, \theta)$ are presented in Table~\ref{tab:helicity_amplitudes} using the Jacob-Wick helicity amplitude conventions~\cite{Jacob_1959} for both photon helicity combinations.
The results in Table~\ref{tab:helicity_amplitudes} are consistent with the results in Ref.~\cite{Nachtmann_2005}.

From Eq.~\eqref{eq:amplitude_helicity}, partial cross sections for $\Lambda = 0,2$ can be calculated separately, and their behavior as a function of center-of-mass energy can be studied.
The expressions for the cross section sum $\sigma_2+\sigma_0$ and the cross section difference $\Delta \sigma = \sigma_2 - \sigma_0$ are given by:
\begin{widetext}
    \begin{equation}
        \begin{aligned}
            \sigma_2 - \sigma_0 &= \frac{e_d^4}{8 \pi s}  \Biggl\{ 38 \sqrt{1 - \frac{4 m_d^2}{s}} - 16 \left( 2 - \frac{5 m_d^2}{s} \right) \tanh ^{-1} \sqrt{1 - \frac{4 m_d^2}{s}} \Biggr\},\\
            \sigma_2 + \sigma_0 &= \frac{2 e_d^4}{8 \pi s} \Bigg\{ \frac{s}{m_d^2} \left( 4 + \frac{3 m_d^2}{s} + \frac{12 m_d^4}{s^2} \right) \sqrt{1 - \frac{4 m_d^2}{s}} - \frac{24 m_d^2}{s} \left( 1 - \frac{2 m_d^2}{s} \right) \tanh ^{-1} \sqrt{1 - \frac{4 m_d^2}{s}} \Bigg\}.\label{eq:linear_combinations}
        \end{aligned}
    \end{equation}  
\end{widetext}
It is shown in Fig.~\ref{fig:s2_s0} that at low energies the helicity-$2$ cross section $\sigma_2$ dominates, while at high energies the helicity-$0$ cross section $\sigma_0$ is dominant.

In order to calculate the two-photon production of tetraquark states, we next consider the overlap of the produced diquark-antidiquark with the final tetraquark.
For two-photon production, charge conjugation invariance only allows $C=+1$ states, and the Landau-Yang theorem prohibits $J=1$ states.
Comparing to the states calculated with the Cornell potential previously, the states that survive are
\begin{equation}
    {}^1 S_0, \ {}^5 S_2, \ {}^3 P_0, \ {}^3 P_2, \ {}^1 D_2, \ {}^5 D_0, \ {}^5 D_2, \ {}^5 D_3, \ {}^5 D_4.\nonumber
\end{equation}
The non-relativistic calculation from Section \ref{sec:Model} can be tied to the perturbative amplitude via the convolution integral, analogously as was done in Ref.~\cite{Danilkin_2017} for quarkonium states.
For a state with fixed quantum numbers and photons with fixed helicities, the matrix element is given by
\begin{widetext}
    \begin{equation}
        \begin{aligned}
            \langle \mathbf{q} \lambda_1, -\mathbf{q} \lambda_2 | \hat{M} | n (LS) J M_J \rangle &= \delta_{M_J \Lambda} \frac{1}{(2 \pi)^2} \int_0^\infty \text{d}p \frac{p^2}{2 E(p)} \Tilde{R}_{n(LS)J} (p) \left( \frac{2L+1}{4\pi} \right)^{1/2} \sum_{\lambda_+} \sum_{\lambda_-} \langle 1 \lambda_+,1 -\lambda_- |S \Lambda_d \rangle \\
            & \qquad\qquad\quad \times \langle L 0,S \Lambda_d | J \Lambda_d \rangle \int_{-1}^1 \text{d} \cos \theta \ d_{\Lambda \Lambda_d}^J (\theta) \,h_{\lambda_+ \lambda_-}^{\lambda_1 \lambda_2} (\beta, \theta),
        \end{aligned}
        \label{eq:matrix_element_total}
    \end{equation}
\end{widetext}
with $\beta = p/E(p) = p/\sqrt{p^2 + m_d^2}$, where $d_{\Lambda \Lambda_d}^J (\theta)$ is the Wigner $d$-matrix element, and where $\Lambda_d \equiv \lambda_+-\lambda_-$.
The radial part of the momentum-space wavefunction $\Tilde{R}_{n(LS)J} (p)$ is connected to $R_{n(LS)J} (r)$ via
\begin{equation}
    \Tilde{R}_{N L} (p)= (4 \pi) \int_0^\infty \text{d} r \ r^2 (-i)^{L} j_{L} (pr) R_{N L} (r) \label{eq:transformations},
\end{equation}
where $j_{L}(pr)$ corresponds to the spherical Bessel functions of the first kind.
Note that in Eq.~(\ref{eq:matrix_element_total}), the two-photon state $|\mathbf{q} \lambda_1, -\mathbf{q} \lambda_2 \rangle$ as well as the diquark-antidiquark state 
$| \mathbf{p} \lambda_+ ,-\mathbf{p} \lambda_- \rangle$ are covariantly normalized, whereas for the tetraquark state $| n (LS) J M_J \rangle$ we conveniently use the normalization:
\begin{equation}
\langle n (LS) J M_J | n (LS) J M_J \rangle = 1.
\end{equation}

\begin{table*}
    \caption{Two-photon decay widths (in $\unit{\kilo\eV}$) for the states lying within the first four energy shells of the all-charm tetraquark spectrum using the parameters displayed in Eq.~\eqref{eq:paper_parameters}. The lower two rows give the  contribution to the sum rule Eq.~(\ref{eq:narrow_width}) of the $\Lambda = 2$ transitions as well as the total sum rule contribution (in units $10^{-6}$ GeV$^{-2}$) respectively.}
    \begin{ruledtabular}
        \begin{tabular}{ccccccccc}
         & \multicolumn{2}{c}{$n=1$} & \multicolumn{2}{c}{$n=2$} & \multicolumn{2}{c}{$n=3$} & \multicolumn{2}{c}{$n=4$} \\ \hline
        states & $\Gamma_{\Lambda=0} \ (\unit{\kilo\eV})$ & $\Gamma_{\Lambda=2} \ (\unit{\kilo\eV})$ & $\Gamma_{\Lambda=0} \ (\unit{\kilo\eV})$ & $\Gamma_{\Lambda=2} \ (\unit{\kilo\eV})$ & $\Gamma_{\Lambda=0 } \ (\unit{\kilo\eV})$ & $\Gamma_{\Lambda=2} \ (\unit{\kilo\eV})$ & $\Gamma_{\Lambda=0} \ (\unit{\kilo\eV})$ & $\Gamma_{\Lambda=2} \ (\unit{\kilo\eV})$ \\ \hline
        $^1S_0$ & 84.6                 & 0                    & 34.0                 & 0                    & 26.4                 & 0                    & 24.0                 & 0                    \\
        $^5S_2$ &  0.2                 & 18.8                 & 0.1                  & 8.3                  & $8.7 \times 10^{-2}$ & 6.3                  & $8.2 \times 10^{-2}$ & 5.4                  \\
        $^3P_0$ & 21.1                 & 0                    &  21.5                & 0                    & 21.4                 & 0                    & 21.3                 & 0                    \\
        $^3P_2$ & $2.6 \times 10^{-2}$ & 0                    & $4.0 \times 10^{-2}$ & 0                    & $5.0 \times 10^{-2}$ & 0                    & $6.0 \times 10^{-2}$ & 0                    \\
        $^1D_2$ & $9.1 \times 10^{-3}$ & 4.3                  & $1.9 \times 10^{-2}$ & 1.4                  & $2.9 \times 10^{-2}$ & 1.8                  & $3.9 \times 10^{-2}$ & 1.1                  \\
        $^5D_0$ & 26.6                 & 0                    & 11.7                 & 0                    & 15.6                 & 0                    & 11.6                 & 0                    \\
        $^5D_2$ & $4.8 \times 10^{-3}$ & 10.5                 & $8.6 \times 10^{-3}$ & 4.8                  & $1.2 \times 10^{-2}$ & 6.5                  & $1.5 \times 10^{-2}$ & 4.9                  \\
        $^5D_3$ & 0                    & $2.0 \times 10^{-2}$ & 0                    & $3.5 \times 10^{-2}$ & 0                    & $4.8 \times 10^{-2}$ & 0                    & $6.0 \times 10^{-2}$ \\
        $^5D_4$ & $2.2 \times 10^{-5}$ & $4.9 \times 10^{-4}$ & $5.6 \times 10^{-5}$ & $9.7 \times 10^{-4}$ & $1.0 \times 10^{-4}$ & $1.4 \times 10^{-3}$ & $1.5 \times 10^{-4}$ & $1.9 \times 10^{-3}$ \\ \hline
        sum rule ($10^{-6}$ GeV$^{-2}$)  & \multicolumn{2}{c}{} & \multicolumn{2}{c}{} & \multicolumn{2}{c}{} & \multicolumn{2}{c}{} \\
        $\Lambda=2$  & \multicolumn{2}{c}{101.5} & \multicolumn{2}{c}{35.4} & \multicolumn{2}{c}{30.5} & \multicolumn{2}{c}{21.7} \\
        total  & \multicolumn{2}{c}{12.3} & \multicolumn{2}{c}{1.2} & \multicolumn{2}{c}{2.7} & \multicolumn{2}{c}{-0.7}
        \end{tabular}
        \label{tab:decay_widths}
    \end{ruledtabular}
\end{table*}

Knowledge of the matrix elements allows us to calculate the two-photon decay widths.
Each width has two distinct contributions, one for $\Lambda = 0$ and one for $\Lambda = 2$.
For $T_J$ states with $J=0$, there is only a helicity-$0$ contribution, i.e.
\begin{eqnarray}
    \Gamma_{\Lambda=0}(T_0) &=& \frac{1}{8 \pi } \left|  \langle \mathbf{q} \, 1, -\mathbf{q} \, 1 | \hat{M} | n (LS) 0 0 \rangle \right|^2, \nonumber \\
        \Gamma_{\Lambda=2}(T_0) &=& 0. \label{eq:spin0}
\end{eqnarray}
For $J \geq 2$, both polarizations should be accounted for, yielding
\begin{eqnarray}
        \Gamma_{\Lambda=0}(T_J) &= &\frac{1}{8 \pi (2J+1)} 
        \left|  \langle \mathbf{q} \, 1, -\mathbf{q} \, 1 | \hat{M} | n (LS) J 0 \rangle \right|^2 , \nonumber \\
        \Gamma_{\Lambda=2}(T_J) &= &\frac{1}{8 \pi (2J+1)}
        \left|  \langle \mathbf{q} \, 1, -\mathbf{q} -\!1 | \hat{M} | n (LS) J 2 \rangle \right|^2 , \nonumber \\
    \label{eq:spin2}
\end{eqnarray}

The calculation detailed above was repeated for every allowed state.
The results for the states computed using the parameters of Eq.~\eqref{eq:paper_parameters} are shown in Table~\ref{tab:decay_widths}.
The first four energy shells have been included, and both helicity contributions are displayed.
The explicit formulas for the states with the largest two-photon widths are given in Appendix~\ref{ap:matrixelements}.

\section{\label{sec:Results and discussion} Results and Discussion}

For forward light-by-light scattering, several sum rules of experimentally measured quantities have been established~\cite{Pascalutsa:2010sj,Pascalutsa:2012pr}. 
In this work we study the implication of the helicity difference sum rule for two real photons, that is:
\begin{equation}
    \int_{s_0}^\infty \frac{ds}{s} [\sigma_2 (s) -\sigma_0 (s)]=0.\label{eq:sum_rule}
\end{equation}
It was first inferred in Refs.~\cite{Gerasimov_1975,Brodsky:1995fj} from the Gerasimov-Drell-Hearn (GDH) sum rule \cite{Gerasimov_1965,Drell_1966}.
The superconvergence relation of Eq.~\eqref{eq:sum_rule} is one of several model-independent light-by-light sum rules based on such general principles as gauge invariance, unitarity, analyticity, and a good high-energy behavior~\cite{Pascalutsa:2012pr}. They have been shown to hold explicitly in renormalizable perturbative quantum field theories. An explicit non-perturbative study has also been performed within the context of $\phi^4$ theory, for a particular resummation of bubble graphs to all orders, which was also found to be consistent with the photon helicity sum rule~\cite{Pauk:2013hxa}. The sum rules have also been studied non-perturbatively using empirical data of transition form factors for light-quark mesons~\cite{Danilkin:2016hnh}, where it was shown that they are predominantly saturated by a few lowest-lying mesons. It has also been studied in the two-photon decays of charmonium states in~\cite{Danilkin_2017}, and in the radiative transitions of bottomonia in~\cite{Ananyev:2020uve}. In the latter work, the helicity sum rule was tested in a Cornell potential framework, and the helicity-2 contribution was found to be canceled between 90-95\% by the helicity-0 contribution. Furthermore, it was observed that the sum rule is satisfied separately for the lowest three shells within experimental errors. 

Being a model-independent result, a consistent model is expected to satisfy the sum rule of Eq.~\eqref{eq:sum_rule}. In the following, we test it for the tetraquark wave function model of Section \ref{sec:Model}.
Within this model, the helicity amplitudes for the $\gamma\gamma \to \text{tetraquark}$ process are obtained as a convolution between the helicity amplitudes for $\gamma\gamma \to d \bar{d}$, with $d$ an axial-vector diquark, and the wave function of the tetraquark as a bound state of diquark-antidiquark, as Eq.~(\ref{eq:matrix_element_total}) shows.
The sum rule Eq.~\eqref{eq:sum_rule} requires that the difference between both energy-weighted helicity cross sections cancels when integrated.
For the $\gamma\gamma \to d \bar{d}$ process, the helicity-2 contribution dominates at lower energies, whereas the helicity-0 contribution dominates at higher energies; see Fig.~\ref{fig:s2_s0} in Appendix~\ref{ap:two_photon}. Using the expression of 
Eq.~(\ref{eq:linear_combinations}) for the helicity difference cross section $\Delta \sigma$ for the $\gamma \gamma$ production of pointlike spin-$1$ diquark and antidiquark states, it is readily verifiable that this sum rule is exactly satisfied.

\begin{figure}
    \includegraphics[width=0.48\textwidth]{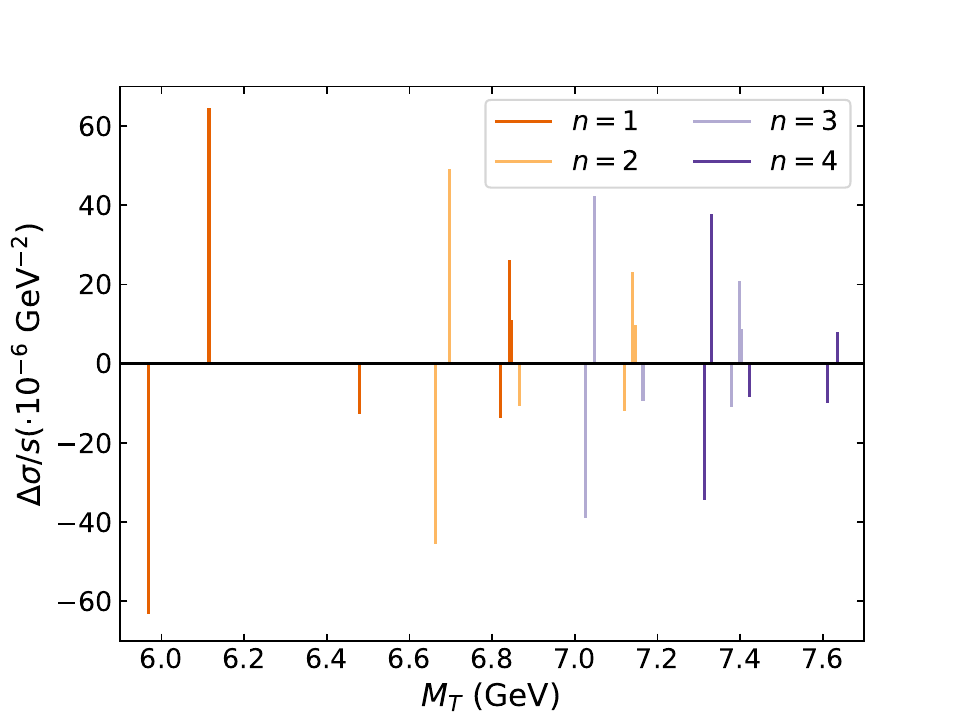}
    \includegraphics[width=0.48\textwidth]{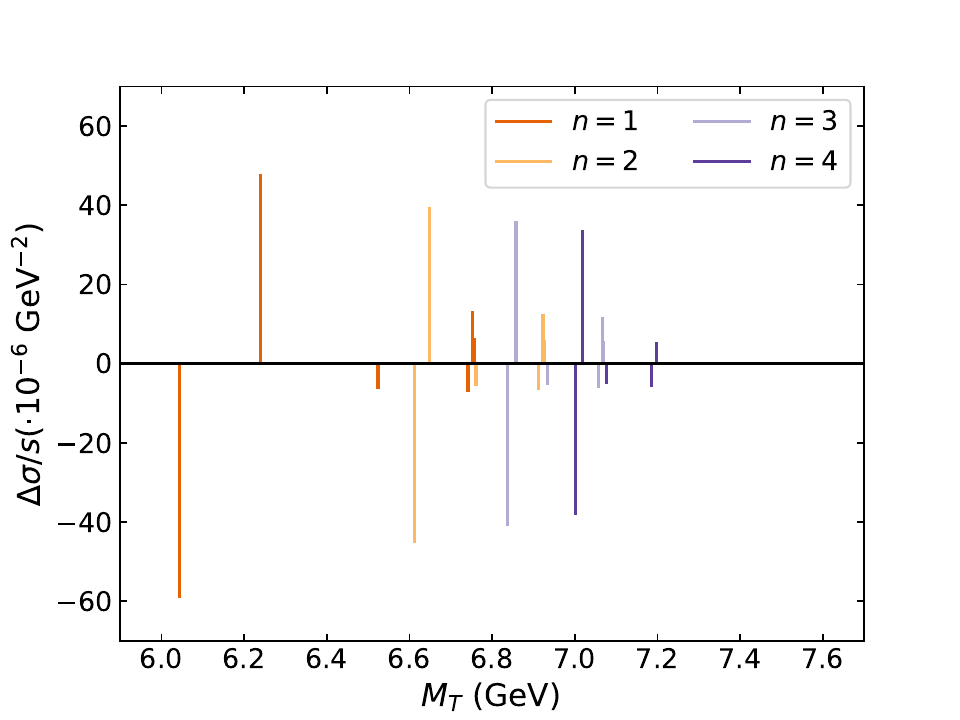}
    \caption{Sum rule contribution of the tetraquark states for different shells $n$ against their mass $M_T$. The model contributions have been calculated using the parameters of Eq.~\eqref{eq:paper_parameters} (upper panel) and using the fit parameters of Eq.~\eqref{eq:fit_parameters} (lower panel).}
    \label{fig:barcode_paper}
\end{figure}

We next consider the contribution of each tetraquark state to the sum rule of Eq.~\eqref{eq:sum_rule}. Following Ref.~\cite{Danilkin_2017}, based on the narrow-width approximation, the sum rule integral for tetraquarks can be expressed as:
\begin{equation}
    \begin{aligned}
        \int_{s_0}^\infty \frac{ds}{s} \Delta \sigma =& -16\pi^2 \Bigg( \sum_{T_0} \frac{\Gamma_{\Lambda = 0} (T_0)}{M_{T_0}^3} \\
        & + \sum_{ T_J \atop J \geq 2} (2J+1) \frac{-\Gamma_{\Lambda=2} (T_J)+\Gamma_{\Lambda=0} (T_J)}{M_{T_J}^3} \Bigg). \label{eq:narrow_width}
    \end{aligned}
\end{equation}
Using the two-photon widths of Table~\ref{tab:decay_widths}, Eq.~\eqref{eq:narrow_width} allows us to calculate the sum-rule contribution of each tetraquark state. 
The resulting components are shown in Figure~\ref{fig:barcode_paper} (upper panel) compared to their respective masses.
Furthermore, the $\Lambda =2$ and total sum rule contributions of each $n$-shell in the potential model are shown in the two last rows of Table~\ref{tab:decay_widths}. 
One notices from Table~\ref{tab:decay_widths} that the main contributions to the sum rule integral of Eq.\eqref{eq:narrow_width} 
come from the states $^1S_0$, $^3P_0$, $^5D_0$ for $\Lambda=0$, 
and from the states $^5S_2$, $^1D_2$, $^5D_2$ for $\Lambda=2$. 
The helicity-0 contribution to the sum rule compensates the helicity-2 contribution to around $90\%$ for each of the energy shells. Thus, one observes that the sum rule is already satisfied relatively well in the potential model and to a good approximation even within each $n$-shell.

Assuming that the integrated strength in the physical spectrum satisfies the sum rule of Eq.~\eqref{eq:sum_rule}, the mass spectrum in the tetraquark model can be fitted to the experimentally verified states in such a way as to further improve consistency with the helicity sum rule. Due to the small number of observed states so far, 
we vary a limited number of parameters as a first step in the present work. 
Of the parameters entering Eq.~\eqref{eq:paper_parameters}, $\alpha_s$ is an inherent feature of strong interactions and should therefore remain the same, while $\sigma$ is an artifact of the potential model approach that has a relatively small effect on the sum rule behavior.
On the other hand, the value of the linear confinement slope parameter $b$ for diquarks is not necessarily the same as its value for quarks. 
Increasing the confinement slope parameter corresponds to stretching the spectrum.  
Furthermore, for the dicharm mass parameter $m_d$, different phenomenological approaches end up with a wide range of values between $2.770$ - $\SI{4.015}{\giga\eV}$. 
Varying the value of $m_d$ corresponds to a shift of the spectrum. 

In a first step to further improve the consistency with the helicity sum rule, we thus vary both $b$ and $m_d$, keeping the di-quark mass $m_d$ within the interval mentioned above. 
Meanwhile, the confinement parameter $b$ is kept positive and allowed to vary from $0-\SI{0.5}{\giga\eV\tothe{2}}$.
In order to further alter the relative weights of the helicity-$2$ versus helicity-$0$ contributions to the discussed sum rule, we also consider varying the spin-spin interaction. The latter affects the energy splitting between states with the same value of $L$.
For this purpose, we introduce a rescaling factor $c_s$ in Eq.~(\ref{eq:potential_specific}) as
\begin{equation}
    \mathcal{V}_{SS} \to c_s \mathcal{V}_{SS}.
\end{equation}
For $c_s = 1$ the potential is identical to that of Eq.~\eqref{eq:generalv}.
For $c_s >1$ the splitting between states with the same value of $L$ and different values of total spin $S$ increases, while for $c_s <1$ it decreases and would be even reversed for negative values.

The potential model is then fitted to the potential tetraquark states reported in Ref.~\cite{CMS_2024}. 
For that purpose,
the results of Ref.~\cite{Zhang:2020hoh} were considered, in which the authors suggested that the observed resonances in di-$J/\psi$ 
experiments correspond to spin-$2$ states. The claim is
based on a perturbative QCD analysis, in which spin-$2$ 
states lead to hadroproduction cross sections more than
an order of magnitude larger than their spin-$0$ counterparts. 
Combining this with the most plausible assumption that the lowest lying observed states correspond with S-wave orbital states, leads to the following identification with the states reported in Table~\ref{tab:CMS_masses} 
\begin{equation}
    \begin{aligned}
        M(2\, ^5 S_2) & \approx M_1,\\
        M(3\, ^5 S_2) & \approx M_2.
        \label{eq:mass_id}
    \end{aligned}
\end{equation}

At this point, we refrain from including the third state observed in di-$J/\psi$ decays, around 7.1~GeV, in our fit, since there are many tetraquark states in that energy range. Our initial fit serves as a proof of principle, which can be improved upon once the quantum numbers of the so-far observed states are established and possibly more states will be discovered, which then will allow testing different scenarios within the tetraquark potential model.      
Changing the parameters $b$ and $m_d$ in the altered potential within the intervals mentioned above, while assigning $c_s$ the value that minimizes the total contribution of the $n = 1, 2, 3$ shells to the sum rule, thus leads to a new mass spectrum. 
In fitting the model to the experimental states according to Eq.~(\ref{eq:mass_id}), the weighted least squares method was used. Consequently, asymmetric uncertainties from the experimental results were considered separately. The resulting fit parameters are:
\begin{equation}
    \begin{gathered}
        m_d =  3.25 \pm \SI{0.06}{\giga\eV},\\
        b= 0.058 \pm \SI{0.027}{\giga\eV\tothe{2}},\\
        c_s = 1.52.
        \label{eq:fit_parameters}
    \end{gathered}
\end{equation}

For an initial estimation of the errors, we created four data sets by fitting the model using the higher and lower limits for the masses of Table~\ref{tab:CMS_masses}.
The standard deviation of the resulting parameters is then shown in Eq.~\eqref{eq:fit_parameters} as a rough estimate of the parameter errors.
In the future, a more refined error estimation could be obtained using a bootstrap method.

\begin{figure*}
    \centering
    \includegraphics[width=0.75\textwidth]{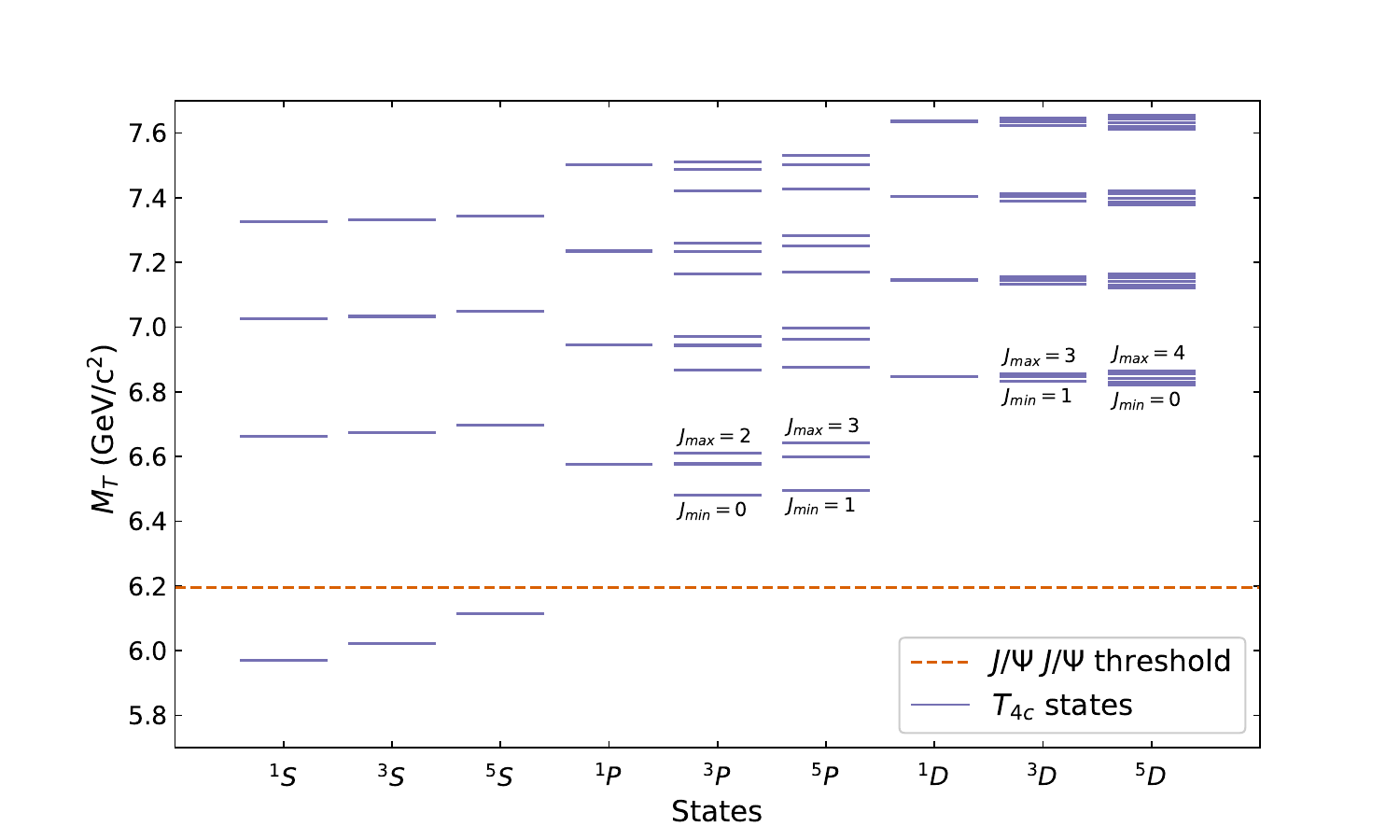}
    \caption{All-charm tetraquark mass spectrum with $1\, ^3 S_1$ axial-vector diquarks for the first four energy shells and parameters from Eq.~\eqref{eq:paper_parameters}.The blue lines represent individual states and the dashed red line is the physical di-$J/\Psi$ threshold. When multiple $J$-values are possible, the minimum ($J_{min}$) and maximum ($J_{max}$) allowed $J$-values are displayed.}
    \label{fig:tetraquark_mass_spectrum_paper}
\end{figure*}

\begin{figure*}
    \centering
    \includegraphics[width=0.75\textwidth]{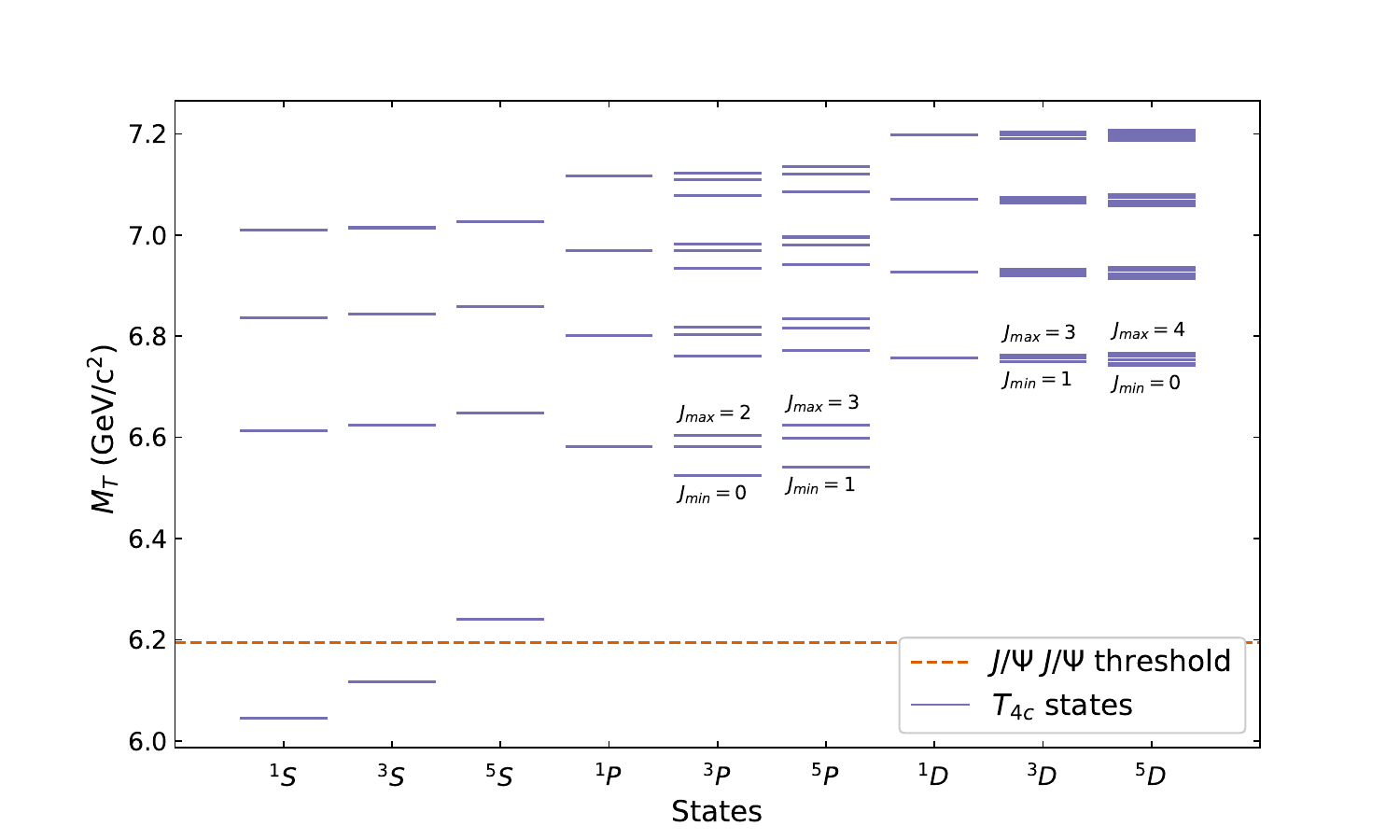}
    \caption{Same as Fig.~\ref{fig:tetraquark_mass_spectrum_paper} when the model parameters are given by Eq.~\eqref{eq:fit_parameters}, which were obtained by the identification of Eq.~(\ref{eq:mass_id}) with the two lowest observed states given in Table~\ref{tab:CMS_masses}, and by requiring consistency with the LbL sum rule of Eq.~\eqref{eq:sum_rule}.}
    \label{fig:tetraquark_mass_spectrum_fit}
\end{figure*}

The same method as in Section~\ref{sec:Model} was applied to solve the Schr\"odinger equation with the altered potential to obtain the all-charm tetraquark spectrum.
The resulting tetraquark mass states are presented in Table~\ref{tab:masses_dif}. 
A side-by-side comparison of the tetraquark spectra produced using the parameters of Eq.~\eqref{eq:paper_parameters} and of Eq.~\eqref{eq:fit_parameters} is shown in Figs.~\ref{fig:tetraquark_mass_spectrum_paper} and \ref{fig:tetraquark_mass_spectrum_fit}, respectively. 
In both cases, states can be observed lying near the di-$J/\Psi$ threshold.

Due to the larger strength of the spin-spin interaction term, the mass gaps between states with the same $L$ of Fig.~\ref{fig:tetraquark_mass_spectrum_fit} are noticeably larger than those of Fig.~\ref{fig:tetraquark_mass_spectrum_paper}. 
Moreover, due to the smaller value of the confinement parameter $b$, the newly fitted mass spectrum is also compressed relative to the original, covering masses from $6.04-\SI{7.21}{\giga\eV}$ compared to the spectrum of Fig.~\ref{fig:tetraquark_mass_spectrum_paper}.

\begin{table}
    \caption{All-charm tetraquark masses with $1\, ^3 S_1$ diquarks for the parameters given in Eq.~\eqref{eq:fit_parameters}. The error is present on the last digit.}
    \label{tab:masses_dif}
    \begin{ruledtabular}
        \begin{tabular}{ccccc}
            &  \multicolumn{4}{c}{$M_T \ (\SI{}{\giga\eV})$} \\ \hline
            & & & & \\[-0.9em]
            ${}^{2S+1} {L}_{J}$ & $n=1$ & $n=2$ & $n=3$ & $n=4$ \\ 
            & & & & \\[-0.9em]
            \hline
            ${}^1 S_0$ & 6.04 & 6.61 & 6.84 & 7.01 \\
            ${}^3 S_1$ & 6.12 & 6.62 & 6.84 & 7.01 \\
            ${}^5 S_2$ & 6.24 & 6.65 & 6.86 & 7.03 \\
            ${}^1 P_1$ & 6.58 & 6.80 & 6.97 & 7.12 \\
            ${}^3 P_0$ & 6.52 & 6.76 & 6.93 & 7.08 \\
            ${}^3 P_1$ & 6.58 & 6.80 & 6.97 & 7.11 \\
            ${}^3 P_2$ & 6.60 & 6.82 & 6.98 & 7.12 \\
            ${}^5 P_1$ & 6.54 & 6.77 & 6.94 & 7.09 \\
            ${}^5 P_2$ & 6.60 & 6.82 & 6.98 & 7.12 \\
            ${}^5 P_3$ & 6.62 & 6.83 & 7.00 & 7.13 \\
            ${}^1 D_2$ & 6.76 & 6.93 & 7.07 & 7.20 \\
            ${}^3 D_1$ & 6.75 & 6.92 & 7.06 & 7.19 \\
            ${}^3 D_2$ & 6.76 & 6.93 & 7.07 & 7.20 \\
            ${}^3 D_3$ & 6.76 & 6.93 & 7.07 & 7.20 \\
            ${}^5 D_0$ & 6.74 & 6.91 & 7.06 & 7.19 \\
            ${}^5 D_1$ & 6.75 & 6.92 & 7.06 & 7.19 \\
            ${}^5 D_2$ & 6.75 & 6.92 & 7.07 & 7.20 \\
            ${}^5 D_3$ & 6.76 & 6.93 & 7.07 & 7.20 \\
            ${}^5 D_4$ & 6.77 & 6.94 & 7.08 & 7.21 \\
        \end{tabular}
    \end{ruledtabular}
\end{table}

Following the procedure detailed in Section~\ref{sec:Two-photon} we next obtain the two-photon widths for the states of this fit.
The resulting widths, as well as the sum rule contribution of each energy shell are presented in Table~\ref{tab:decay_widths_dif}.
The sum rule behavior for the new fit is presented in Fig.~\ref{fig:barcode_paper} (lower panel) and shown alongside the calculation using the original parameters (upper panel).
One notices that the fit has evidently changed the contributions compared to their strength using the initial parameters.
The contributions for both $\Lambda=0$ and $\Lambda=2$ are smaller than before and have shifted on the mass axis. Summing over the first three energy shells, Table~\ref{tab:decay_widths_dif} shows that the helicity-0 contribution to the sum rule compensates the helicity-2 contribution to around 99.9\%.  
Consequently, the results coming from the model are now consistent with what is expected from the model-independent sum rule for the first three energy shells, where the relevant states lie.

\begin{table*}
    \caption{Two-photon decay widths (in $\unit{\kilo\eV}$) for the states lying within the first four energy shells of the all-charm tetraquark spectrum using the parameters displayed in Eq.~\eqref{eq:fit_parameters}. The lower two rows give the  contribution to the sum rule Eq.~(\ref{eq:narrow_width}) of the $\Lambda = 2$ transitions as well as the total sum rule contribution (in units $10^{-6}$ GeV$^{-2}$) respectively.}
    \begin{ruledtabular}
        \begin{tabular}{ccccccccc}
         & \multicolumn{2}{c}{$n=1$} & \multicolumn{2}{c}{$n=2$} & \multicolumn{2}{c}{$n=3$} & \multicolumn{2}{c}{$n=4$} \\ \hline
        states & $\Gamma_{\Lambda=0} \ (\unit{\kilo\eV})$ & $\Gamma_{\Lambda=2} \ (\unit{\kilo\eV})$ & $\Gamma_{\Lambda=0} \ (\unit{\kilo\eV})$ & $\Gamma_{\Lambda=2} \ (\unit{\kilo\eV})$ & $\Gamma_{\Lambda=0 } \ (\unit{\kilo\eV})$ & $\Gamma_{\Lambda=2} \ (\unit{\kilo\eV})$ & $\Gamma_{\Lambda=0} \ (\unit{\kilo\eV})$ & $\Gamma_{\Lambda=2} \ (\unit{\kilo\eV})$ \\ \hline
        $^1S_0$ &  82.1                & 0                    & 20.7                 & 0                    & 14.5                 & 0                    & 12.4                 & 0                    \\
        $^5S_2$ &  0.1                 & 14.7                 & $5.5 \times 10^{-2}$ &  5.1                 & $4.2 \times 10^{-2}$ & 3.6                  & $3.7 \times 10^{-2}$ & 3.0                  \\
        $^3P_0$ & 10.5                 & 0                    &  9.6                 & 0                    &  9.1                 & 0                    &  8.9                 & 0                    \\
        $^3P_2$ & $1.1 \times 10^{-2}$ & 0                    & $1.3 \times 10^{-2}$ & 0                    & $1.5 \times 10^{-2}$ & 0                    & $1.6 \times 10^{-2}$ & 0                    \\
        $^1D_2$ & $1.9 \times 10^{-3}$ & 2.4                  & $3.7 \times 10^{-3}$ & 0.9                  & $5.3 \times 10^{-3}$ & 1.1                  & $6.9 \times 10^{-3}$ & 0.7                  \\
        $^5D_0$ & 13.1                 & 0                    &  5.7                 & 0                    &  7.3                 & 0                    &  5.4                 & 0                    \\
        $^5D_2$ & $1.1 \times 10^{-3}$ &  5.1                 & $1.8 \times 10^{-3}$ & 2.3                  & $2.4 \times 10^{-3}$ & 2.9                  & $2.9 \times 10^{-3}$ & 2.2                  \\
        $^5D_3$ & 0                    & $4.2 \times 10^{-3}$ & 0                    & $7.1 \times 10^{-3}$ & 0                    & $9.6 \times 10^{-3}$ & 0                    & $1.2 \times 10^{-2}$ \\
        $^5D_4$ & $3.1 \times 10^{-6}$ & $9.4 \times 10^{-5}$ & $6.9 \times 10^{-6}$ & $1.7 \times 10^{-4}$ & $1.1 \times 10^{-5}$ & $2.4 \times 10^{-4}$ & $1.6 \times 10^{-5}$ & $3.1 \times 10^{-4}$ \\ \hline
        sum rule ($10^{-6}$ GeV$^{-2}$)  & \multicolumn{2}{c}{} & \multicolumn{2}{c}{} & \multicolumn{2}{c}{} & \multicolumn{2}{c}{} \\
        $\Lambda=2$  & \multicolumn{2}{c}{66.9} & \multicolumn{2}{c}{21.3} & \multicolumn{2}{c}{17.8} & \multicolumn{2}{c}{13.1} \\
        total  & \multicolumn{2}{c}{-5.0} & \multicolumn{2}{c}{2.2} & \multicolumn{2}{c}{2.9} & \multicolumn{2}{c}{1.0}
        \end{tabular}
    \label{tab:decay_widths_dif}
    \end{ruledtabular}
\end{table*}

Ref.~\cite{Biloshytskyi_2022} has explored the excess seen in the LbL scattering data observed by the ATLAS Collaboration~\cite{ATLAS_2017} in the 5 - 10~GeV two-photon mass range. The excess observed was fitted to the required two-photon width of the observed resonance $X(6900)$, assuming that the entire contribution comes from the $X(6900)$ state, in two scenarios for the mass and width of this state. 

In Table \ref{tab:bil_results} we show the total two-photon decay widths deduced in~\cite{Biloshytskyi_2022} needed to make the Standard Model prediction consistent with the ATLAS results~\cite{ATLAS_2017} in the $5 - \SI{10}{\giga\eV}$ energy region in both scenarios (labeled ``interference" and ``no-interference"). 
To compare this result with our tetraquark potential model, we make 
an initial comparison using a narrow resonance model, in which the $\gamma \gamma \to T_J$ cross section for production of a tetraquark state with spin $J$ is given by~\cite{Pascalutsa:2012pr}
\begin{equation}
\sigma(\gamma \gamma \to T_J) = 8 \pi^2 \delta(s - M_T^2) \frac{(2J+1) \Gamma_{\gamma \gamma}(T_J)}{M_T},
\end{equation}
where the total two-photon decay width $\Gamma_{\gamma \gamma}(T_J)$ of the tetraquark state $T_J$ is given by
\begin{equation}
\Gamma_{\gamma \gamma}(T_J) = \Gamma_{\Lambda = 0}(T_J) + \Gamma_{\Lambda = 2}(T_J).
\end{equation}
To compare with the result of ~\cite{Biloshytskyi_2022}, where it was assumed that the entire excess observed in the ATLAS LbL scattering data in the energy interval $5~\mathrm{GeV} < \sqrt{s} < 10~\mathrm{GeV}$ is attributed to one exotic spin-0 state $X(6900)$, we consider the integrated two-photon fusion cross section over the interval $[s_L, s_U]$:
\begin{equation}
I \equiv \int_{s_L}^{s_U} ds \, \sigma(\gamma \gamma \to T_J) = 8 \pi^2 \sum_{T_J} \frac{(2J+1) \Gamma_{\gamma \gamma}(T_J)}{M_T},
\label{eq:intcross}
\end{equation}
where the sum extends over the tetraquark states in the considered interval $[s_L, s_U]$. Identifying the three observed states of Table~\ref{tab:CMS_masses} with the $n \, ^5S_2$ states for $n = 2, 3, 4$ in our tetraquark model, we show the resulting integrated two-photon fusion cross section strength in Table \ref{tab:bil_results}.  
\begin{table}[h]
    \centering
    \caption{Two-photon decay width of $X(6900)$ from Ref.~\cite{Biloshytskyi_2022} as well as the resulting integrated strength $I$ of Eq.~(\ref{eq:intcross}), compared with its value for the sum of the $2\, ^5S_2 + 3\, ^5S_2 + 4\, ^5S_2$ states using the parameters of Eq.~\eqref{eq:fit_parameters}.}
    \begin{tabular}{ccc}
    \hline\hline
         & $\Gamma_{\gamma\gamma} (\unit{\kilo\eV})$ & \quad $I$ (in $10^{-4}$) \\ 
         \hline
         & & \\[-0.9em]
        $X(6900)$ interference~\cite{Biloshytskyi_2022} & $67^{+15}_{-19}$ & $7.7 \pm 2.0$ \\
        $X(6900)$ no-interference~\cite{Biloshytskyi_2022} & $45^{+11}_{-14}$ & $5.2 \pm 1.4$ \\
        $2\, ^5S_2 + 3\, ^5S_2 + 4\, ^5S_2$ using Eq.~\eqref{eq:fit_parameters} & & 6.8 \\
        \hline\hline
    \end{tabular}
    \label{tab:bil_results}
\end{table}
One notices from Table~\ref{tab:bil_results} that the tetraquark model with the fit parameters according to Eq.~(\ref{eq:fit_parameters}) gives a contribution for the two-photon production strength of the $2\, ^5S_2 + 3\, ^5S_2 + 4\, ^5S_2$ tetraquark states that lies between both scenarios deduced in Ref.~\cite{Biloshytskyi_2022}. 
A more quantitative comparison with the ATLAS data will require one to directly compare the fiducial cross section for the $\gamma \gamma \to \gamma \gamma$ process including the resonance contributions. 
The tetraquark model presented in this work presents a framework to perform such a quantitative study in the future.

\section{Conclusions and outlook}
\label{sec:Conclusion}

In this work, two-photon widths for all-charm tetraquark states were calculated in a potential framework to model the interaction between two diquarks. We followed and expanded on the description
of the Cornell potential model to calculate the
predicted all-charm tetraquark spectrum, including the
D states as well. The strongest attraction occurs between a color antitriplet $cc$ diquark and a color triplet $\bar c \bar c$ antidiquark, both of which have axial-vector quantum numbers.  

Using the tetraquark bound state wave function, we then calculated the production rate for a tetraquark in the two-photon fusion process or, equivalently, the decay width of the tetraquark state to two photons. The matrix element for the two-photon to diquark-antidiquark
transition was calculated with perturbative methods, by using the same vertices that appear in the Standard Model calculation of the $\gamma \gamma \to W^+ W^-$ process. The non-minimal terms appearing in the corresponding vertices were found to be required for exactly satisfying a model-independent light-by-light sum rule for the difference of helicity-0 and helicity-2 two-photon fusion cross sections. 
The two-photon decay widths of the tetraquark states were then obtained as a convolution integral between the tetraquark wave function as a bound state of diquark-antidiquark, and the elementary amplitude for the diquark-antidiquark to two-photon process.  
The dominant two-photon decay widths were found to result from the states $^1S_0$, $^3P_0$, $^5D_0$ for two-photon helicity-0, 
and from the states $^5S_2$, $^1D_2$, $^5D_2$ for two-photon helicity-2. Using Cornell potential parameters from the literature, it was found that the helicity-0 sum-rule contribution compensates
the helicity-2 sum-rule contribution to around 90\% for each of the
energy shells, characterized by the principal quantum number $n$. Thus, one observes that the model-independent 
helicity sum rule is already satisfied relatively well within the potential model framework and to a good approximation even within each energy shell.

We then performed a new fit of the potential model as a first step to further improve the consistency with the photon-helicity sum rule. The diquark mass and the confinement slope parameter were adjusted by identifying the S-wave orbital tetraquark states $2\,  ^5S_2$ and $3 \, ^5S_2$ with the two lowest observed states. By also varying the spin-spin interaction parameter in the fit, it was found that when summing over the first three energy shells, the helicity-0 contribution to the sum rule compensates the
helicity-2 contribution to around 99.9\%. The resulting slightly modified tetraquark spectrum as well as two-photon decay widths were presented. 
Finally, using the calculated two-photon widths, we made an initial exploration of the excess seen in the light-by-light scattering data observed by the ATLAS Collaboration in the 5 - 10 GeV two-photon mass range. 
The two-photon production strength of the $2\, ^5S_2 + 3\, ^5S_2 + 4\, ^5S_2$ tetraquark states was found to be between both empirical scenarios, deduced in an earlier interpretation of the ATLAS data. 
 
A limiting factor of the initial fit performed in this work was the limited experimental information on potential tetraquark states.
Analyses of additional data from the LHCb, ATLAS, and CMS Collaborations may add potential states to the existing picture and help refine the fit that was performed. Moreover, the limited statistics of LbL scattering will be further improved with higher statistics Run-3 and Run-4 data at the LHC,  providing a clearer picture. Such higher statistics will allow for a narrower energy binning and may lead to a direct observation of the all-charm tetraquark states in the light-by-light process.

For further studies, once more data becomes available, the model employed here can also be refined and enhanced.
Supplementary modifications to the potential used in this work may be implemented to better account for new states.
Model-independent predictions, such as the sum rule that we explored in the present work, may provide a powerful constraint in such model building.

\begin{acknowledgments}

We are very grateful to the anonymous referee for suggesting a check of our helicity amplitudes which allowed us to correct a mistake in the phase convention used in an earlier version of this draft. 

This work was supported by the Deutsche Forschungsgemeinschaft (DFG, German Research Foundation), in part through the Research Unit (Photon-photon interactions in the Standard Model and beyond, Projektnummer 458854507—FOR 5327), and in part through the Cluster of Excellence (Precision Physics, Fundamental Interactions, and Structure of Matter) (PRISMA$^+$ EXC 2118/1) within the German Excellence Strategy (Project ID 39083149).

\end{acknowledgments}

\appendix

\section{Tensor factor calculations\label{ap:tensor_factor}}

The calculation of the tensor factor that produces Table~\ref{tab:tensor_factor} is similar for every combination of $(L,S,J)$.
Displaying the computation for one of these factor is sufficient to understand how to replicate the process.
As in \cite{Debastiani_2019}, only the factors for the states with the maximum projection $M_J$ were calculated.

One of the simplest $D$ states is $(2,1,3)$ whose calculation of the tensor factor is shown here.
In bracket notation it is $| J, M_J \rangle_J = |3,3\rangle_J$, where the subscript $J$ indicates that one works in the $J$-basis.
However, the tensor operator of Eq.~\eqref{eq:tensor_formula} acts on the diquark basis. 
First, the state was decomposed into an $S$-basis vector, denoted by $| S, M_S\rangle_S$, and a spherical harmonic ($Y^{M_L}_L (\theta, \phi)$) using the Clebsch-Gordan coefficients.
Second, the same was performed changing from the $S$ basis to the individual diquark and antidiquark basis.
For this particular state it yields
\begin{equation}
    \begin{aligned}
        | 3,3 \rangle_{J} =& | 1,1 \rangle_{S} Y^2_2 (\theta, \phi) \\
        =& \frac{1}{\sqrt{2}} \left\{ | 1,1 \rangle_{d} | 1,0 \rangle_{\overline{d}} - | 1,0 \rangle_{d} | 1,1 \rangle_{\overline{d}} \right\} Y^2_2 (\theta, \phi).\label{eq:diquarkbase}
    \end{aligned}
\end{equation}
The matrix element under consideration is given by 
\begin{equation}
    \begin{aligned}
        \langle T_{d \overline{d}} \rangle =  {}_S\langle 1,1| Y^2_2 (\theta, \phi) T_{d \overline{d}} Y^2_2 (\theta, \phi) | 1,1 \rangle_{S}.
    \end{aligned}
    \label{eq:tdecomposition}
\end{equation}
One part includes three spherical harmonics integrated over all solid angles.
Such integrals take the well studied form,
\begin{equation}
    \int \text{d} \Omega \ Y^{M'_L*}_{L'} (\theta, \phi) Y^q_2 (\theta, \phi) Y^{M_L}_L (\theta, \phi)
\end{equation}
which is non-vanishing when the following selection rules are obeyed,
\begin{equation}
    \begin{gathered}
    L'=L,L-2,L+2,\\
    M'_L=M_L+q.\label{eq:selectionrules}
\end{gathered}
\end{equation}
Enforcing Eq.~\eqref{eq:selectionrules} means that the surviving terms should have $M'_L=0$.
According to Eq.~\eqref{eq:tensor_specifics}, only the terms $T_{0}$ and $T'_{0}$ survive.

The other part contains the spin operators that act upon the diquark vectors of Eq.~\eqref{eq:diquarkbase}.
The corresponding operators act according to the usual spin eigenvalue relations, these are
\begin{equation}
    \begin{aligned}
        \mathbf{S}_{dz} | 1 , M_S \rangle_{d} &= M_S | 1 , M_S \rangle_{d}, \\
        \mathbf{S}_{d\pm} | 1 , M_S \rangle_{d} &= \sqrt{2 - M_S (M_S \pm 1)} | 1 , M_S \pm 1 \rangle_{d}, \\
        \mathbf{S}_{\overline{d}z} | 1 , M_S \rangle_{\overline{d}} &= M_S | 1 , M_S \rangle_{\overline{d}}, \\
        \mathbf{S}_{\overline{d}\pm} | 1 , M_S \rangle_{\overline{d}} &= \sqrt{2 - M_S (M_S \pm 1)} | 1 , M_S \pm 1 \rangle_{\overline{d}}, \\
    \end{aligned}
    \end{equation}
which yields the expectation value of the spin part of $T_{0}$ as,
\begin{equation}
   {}_{S}\langle 1,1| \mathbf{S}_{d z} \mathbf{S}_{\overline{d}z} | 1,1 \rangle_{S} = 0.
\end{equation}
On the other hand, for $T'_{0}$ the expectation value of the product of the spin operators is,
\begin{equation}
   {}_{S}\langle 1,1| (\mathbf{S}_{d+} \mathbf{S}_{\overline{d}-} + \mathbf{S}_{d-} \mathbf{S}_{\overline{d}+}   ) | 1,1 \rangle_{S} = -2.
\end{equation}

Both $T_0$ and $T'_0$ contain the same integral over spherical harmonics, given by
\begin{equation}
    \int \text{d} \Omega Y^2_2 (\theta, \phi) Y^0_2 (\theta, \phi) Y^2_2 (\theta, \phi) = - \frac{1}{7} \sqrt{\frac{5}{\pi}}.
\end{equation}
Working out the total factors of both contributions yields
\begin{equation}
    \langle T_{d \overline{d}} \rangle = -\frac{8}{7}.
\end{equation}
Repeating the same procedure for every accessible combination of $(L,S,J)$ yields the factors in Table~\ref{tab:tensor_factor}.

\begin{figure}[t]
    \centering
    \includegraphics[height=5cm]{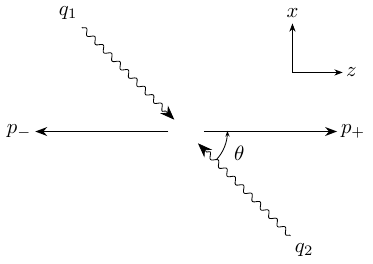}
    \caption{Kinematics of the $\gamma \gamma \to d \bar d$ process, with $d$ the diquark.}
    \label{fig:kinematics}
\end{figure}

\section{$\gamma \gamma \to W^+ W^-$ process}
\label{ap:two_photon}

We consider axial-vector diquarks in this work, with coupling to photons exactly as for the Standard Model $W$-boson. Therefore, the matrix elements for the $\gamma \gamma \to d \bar d$ process have the same form as the $\gamma \gamma \to W^+ W^-$ ones.  
The relevant Feynman rules are \\
\begin{tabular}{ m{2.5cm}  m{5.cm} m{1cm}}
    \parbox[m]{2.5cm}{\includegraphics[width=2.5cm]{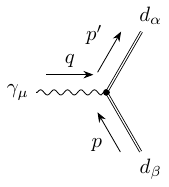}} & \parbox[m]{5.cm}{
    \begin{align*}
        & i e_d \big[ (p+p')^{\mu} g^{\alpha \beta} - g^{\alpha \mu} p'^{\beta} \\
        & - g^{\beta \mu} p^{\alpha} + \left( q^{\alpha} g^{\beta \mu} - q^{\beta} g^{\alpha \mu} \right) \big],
    \end{align*}
    } & $(\text{B}1) \refstepcounter{equation} \label{eq:feynrule1}$ \\
\end{tabular}
where $q=p'-p$, and \\
\begin{tabular}{ m{2.5cm}  m{5.cm} m{1cm}}
    \parbox[m]{2.5cm}{\includegraphics[width=2cm]{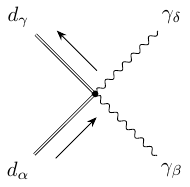}} & \parbox[m]{5.cm}{
    \begin{equation*}
        e_d^2 \left( g^{\alpha \beta} g^{\gamma \delta} + g^{\alpha \delta} g^{\beta \gamma} - 2 g^{\beta \delta} g^{\alpha \gamma} \right).
    \end{equation*}
    } & $(\text{B}2) \refstepcounter{equation} \label{eq:feynrule2}$ \\
\end{tabular}
The unitary gauge was chosen, so the diquark propagator is then \\
\begin{tabular}{ m{2.5cm}  m{5.cm} m{1cm}}
    \parbox[m]{2.5cm}{\includegraphics[width=2.5cm]{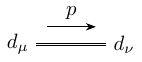}} & \parbox[m]{5.cm}{
    \begin{equation*}
        i \frac{i}{p^2-m_d^2} \left( - g_{\mu \nu} + \frac{p_{\mu} p_{\nu}}{m_d^2} \right).
    \end{equation*}
    } & $(\text{B}3) \refstepcounter{equation} \label{eq:feynrule3}$ \\
\end{tabular}

The kinematics of the $\gamma \gamma \to W^+ W^-$ process is specified in Figure~\ref{fig:kinematics}.
Explicit expressions for the particle polarization vectors are also needed.
In the center-of-mass frame, the photon and $W$-boson  polarization vectors are given as
{\allowdisplaybreaks
\begin{align}
    \nonumber \varepsilon^\mu_{\lambda_1 =\pm1}   &= \varepsilon^\nu_{\lambda_2 =\mp1}= \varepsilon^\mu_{R,L} \\\nonumber
    \varepsilon^\mu_{R,L} &= \frac{1}{\sqrt{2}}(0, \mp\cos \theta, -i, \mp\sin \theta) \\\nonumber
    \varepsilon^\alpha_{\lambda_+ =\pm1} &=\frac{1}{\sqrt{2}}(0,\mp1,-i,0) \\
    \varepsilon^\alpha_{\lambda_+ =0}  &=\frac{\sqrt{s}}{2 m_d}(\beta ,0,0,1) \\
    \nonumber
    \varepsilon^\beta_{\lambda_- =\pm1} &=\frac{1}{\sqrt{2}}(0,\pm1,-i,0) \\ \nonumber
    \varepsilon^\beta_{\lambda_- =0}  &=\frac{\sqrt{s}}{2 m_d}(\beta ,0,0,-1).\\\nonumber \label{eq:polarizations}
\end{align}}
The polarization vectors are chosen to be consistent with the 
Jacob-Wick helicity amplitude conventions~\cite{Jacob_1959}.
The evaluation of the graphs in Figure \ref{fig:two_photon_graphs} can be performed with the assistance of the rules described in Eqs.~(\ref{eq:feynrule1},\ref{eq:feynrule2},\ref{eq:feynrule3}).
The resulting expression from the implementation of the Feynman rules is presented in Eq.~\eqref{eq:amplitude_abc}.
For the helicity-amplitude method, each combination needs to be worked out separately.
After specifying the frame and for $\lambda_1 = \lambda_2 = +1$, the amplitudes become
\begin{widetext}
    \begin{equation}
        \begin{aligned}
            \mathcal{M}_a =& \frac{8 i e_d^2}{s (1 + \beta \cos \theta)} \bigg[ \delta_{\lambda_+ \lambda_-} \left( 1 - \frac{s}{2 m_d^2} \delta_{\lambda_+ =0} \right) \left( - \frac{\sqrt{s}}{2 \sqrt{2}} \beta \sin \theta \right) \left( - \frac{\sqrt{s}}{2 \sqrt{2}} \beta \sin \theta \right) + (q_1 \cdot \varepsilon^*_{\lambda_-}) (q_2 \cdot \varepsilon^*_{\lambda_+}) \\
            & + \frac{\sqrt{s}}{2 \sqrt{2}} \beta \sin \theta (\varepsilon^*_{\lambda_+} \cdot \varepsilon_{L}) (q_2 \cdot \varepsilon^*_{\lambda_-}) +  \frac{\sqrt{s}}{2 \sqrt{2}} \beta \sin \theta  (\varepsilon_{R} \cdot \varepsilon^*_{\lambda_-}) (q_1 \cdot \varepsilon^*_{\lambda_+}) -  \frac{\sqrt{s}}{2 \sqrt{2}} \beta \sin \theta  (\varepsilon_{R} \cdot \varepsilon^*_{\lambda_+}) (q_1 \cdot \varepsilon^*_{\lambda_-})\\
            &- \frac{\sqrt{s}}{2 \sqrt{2}} \beta \sin \theta (\varepsilon_{L} \cdot \varepsilon^*_{\lambda_-}) (q_2 \cdot \varepsilon^*_{\lambda_+}) + (\varepsilon_{L} \cdot \varepsilon^*_{\lambda_+})(\varepsilon_{R} \cdot \varepsilon^*_{\lambda_-}) \frac{s}{8} (3-\beta \cos \theta) \biggr],\\
            \mathcal{M}_b =& \frac{8 i e_d^2}{s (1 - \beta \cos \theta)} \bigg[ \delta_{\lambda_+ \lambda_-} \left( 1 - \frac{s}{2 m_d^2} \delta_{\lambda_+ =0} \right) \left( \frac{\sqrt{s}}{2 \sqrt{2}} \beta \sin \theta \right) \left( \frac{\sqrt{s}}{2 \sqrt{2}} \beta \sin \theta \right) + \delta_{\lambda_2 = +1} (q_1 \cdot \varepsilon^*_{\lambda_+}) (q_2 \cdot \varepsilon^*_{\lambda_-}) \\
            &- \frac{\sqrt{s}}{2 \sqrt{2}} \beta \sin \theta (\varepsilon^*_{\lambda_-} \cdot \varepsilon_{L}) (q_2 \cdot \varepsilon^*_{\lambda_+}) -  \frac{\sqrt{s}}{2 \sqrt{2}} \beta \sin \theta  (\varepsilon_{R} \cdot \varepsilon^*_{\lambda_+}) (q_1 \cdot \varepsilon^*_{\lambda_-}) +  \frac{\sqrt{s}}{2 \sqrt{2}} \beta \sin \theta (\varepsilon_{R} \cdot \varepsilon^*_{\lambda_-}) (q_1 \cdot \varepsilon^*_{\lambda_+})\\
            & + \frac{\sqrt{s}}{2 \sqrt{2}} \beta \sin \theta (\varepsilon_{L} \cdot \varepsilon^*_{\lambda_+}) (q_2 \cdot \varepsilon^*_{\lambda_-}) + (\varepsilon_{R} \cdot \varepsilon^*_{\lambda_+}) (\varepsilon_{L} \cdot \varepsilon^*_{\lambda_-}) \frac{s}{8} (3+\beta \cos \theta) \biggr],\\
            \mathcal{M}_c =& i e_d^2 \biggl[ (\varepsilon_{R} \cdot \varepsilon^*_{\lambda_-}) (\varepsilon_{L} \cdot \varepsilon^*_{\lambda_+}) + (\varepsilon_{R} \cdot \varepsilon^*_{\lambda_+}) (\varepsilon_{L} \cdot \varepsilon^*_{\lambda_-}) -2 \delta_{\lambda_{+} \lambda_{-}} \left( 1-\frac{s}{4 m_d^2} \delta_{\lambda_+ = 0} \right) \delta_{\lambda_2 = +1} \biggr].\label{eq:amplitudel1l2}
        \end{aligned}
    \end{equation}
\end{widetext}
This expression gives the values of the third column of Table~\ref{tab:helicity_amplitudes} when $\lambda_+$ and $\lambda_-$ are specified.
To show this calculation through, we take the example values $\lambda_+ = \lambda_- = +1$.
After some algebra, Eq.~\eqref{eq:amplitudel1l2} takes the following form,

\begin{align}
    \mathcal{M}_a =& \frac{8ie_d^2}{s(1+ \beta \cos \theta)} \bigg[ \sin^2 \theta (1 + \beta)^2 \nonumber \\
    &\qquad\qquad + \frac{1}{4} (1-\cos \theta)^2 (3 - \beta \cos \theta) \bigg], \nonumber
\end{align}
\begin{align}
    \mathcal{M}_b =& \frac{8ie_d^2}{s(1- \beta \cos \theta)} \bigg[ \sin^2 \theta (1 + \beta)^2 \\
    & \qquad\qquad+ \frac{1}{4} (1+\cos \theta)^2 (3 + \beta \cos \theta) \bigg] \nonumber,
\end{align}
\begin{equation}
    \mathcal{M}_c = i e_d^2 \left[ \frac{1}{4} (1+\cos \theta)^2 + \frac{1}{4} (1-\cos \theta)^2 - 2 \right].\nonumber
\end{equation}
Adding these contributions produces the final amplitude,
\begin{equation}
    h^{+1+1}_{+1,+1} (\beta, \theta) = \frac{2 i e_d^2}{1-\beta^2 \cos^2 \theta} (1 + \beta)^2,
\end{equation}
which is presented in Table~\ref{tab:helicity_amplitudes}.
The same operation for different combinations of $\lambda_+$, $\lambda_-$ provides the third column.
Likewise, producing an equation like Eq.~\eqref{eq:amplitudel1l2} for the $\Lambda=2$ contributions and following the same procedure provides the fourth column as well.

\begin{figure}[h!]   
\centering
\includegraphics[width=\columnwidth]{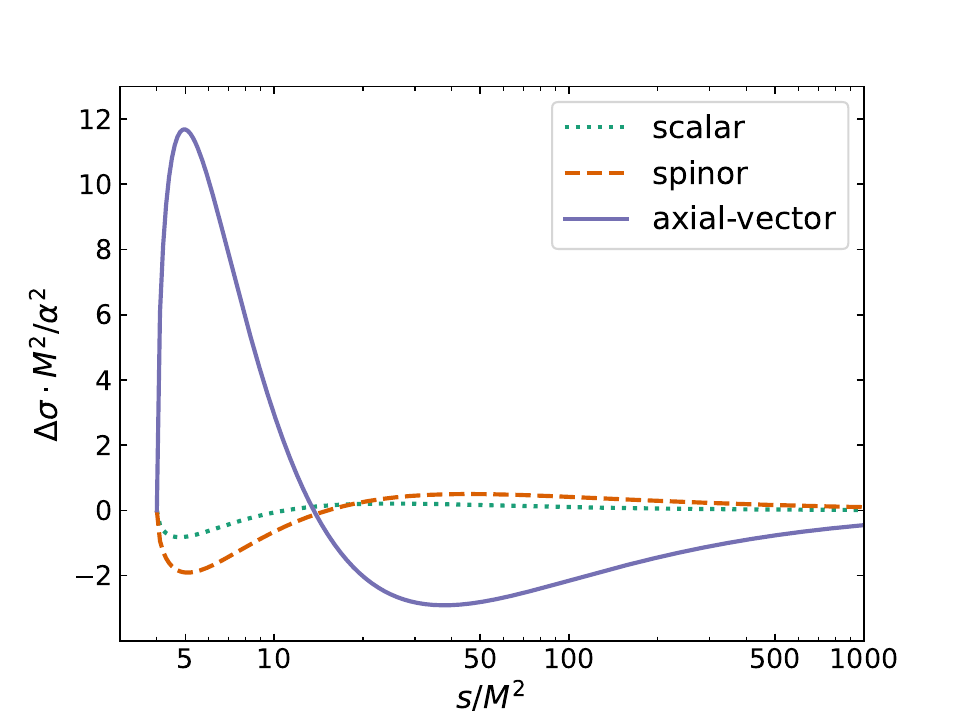}
    \caption{Helicity cross section difference $\Delta \sigma = \sigma_2 - \sigma_0$ for the $\gamma \gamma \to x \bar x$ process as a function of the dimensionless squared c.m. energy $s/M^2$, with $M$ the mass of the produced particle $x$, which is considered with unit charge. Three different cases for the produced particle $x$ are considered: axial-vector (blue solid), spinor (red dashed), and scalar (green dotted).}
    \label{fig:s2_s0}
\end{figure}

The two-photon sum rule of Eq.~\eqref{eq:sum_rule} should be valid for any theory that satisfies tree-level unitarity. 
For the scalar and spinor cases, this was demonstrated in \cite{Pascalutsa:2010sj,Pascalutsa:2012pr}. 
Figure~\ref{fig:s2_s0} shows the c.m. energy behavior of the sum-rule integrand for the aforementioned cases of two-photon production of a pair of spin-0, spin-1/2, and spin-1 particles with all empirical parameters set to one.

One notices from Fig.~\ref{fig:s2_s0} that the spin-1 case differs qualitatively compared to the spinor and scalar cases.  
The overall magnitude is larger by a significant amount.
Additionally, in the low-energy region the photon helicity-2 cross section dominates as compared to the spinor and scalar cases, for which the helicity-0 cross section dominates at low energies. 
The situation is reversed at higher energies in such a way as to guarantee the cancellation needed for the validity of the sum rule.

\section{Tetraquark $n (LS)JM_J \to \gamma \gamma $ matrix elements
\label{ap:matrixelements}}

The explicit formulas for the $n (LS)JM_J \to \gamma \gamma $ matrix elements for the tetraquark states with the largest two-photon decay widths, which are used for the numerical estimates in this paper, follow from Eq.~(\ref{eq:matrix_element_total}). For total photon helicity-0 they are given by: 
\begin{widetext}
\allowdisplaybreaks
\begin{eqnarray}
\langle \mathbf{q} +1, -\mathbf{q} +1 | \hat{M} | n (00) 00 \rangle
&=& i \left(\frac{4}{3}\right)^2 \frac{\alpha}{2 \pi^{3/2}} \int_0^\infty \frac{\text{d}p \, p^2}{E(p)} \Tilde{R}_{n(00)0} (p) \, \frac{2}{\sqrt{3}} \frac{(1 + 3 \beta^2)}{\beta} Q_0\left(\frac{1}{\beta}\right), \label{eq:matrix_elements_detail1} \\
\langle \mathbf{q} +1, -\mathbf{q} +1 | \hat{M} | n (02) 20 \rangle 
&=& i \left(\frac{4}{3}\right)^2 \frac{\alpha}{2 \pi^{3/2}} \int_0^\infty \frac{\text{d}p \, p^2}{E(p)} \Tilde{R}_{n(02)2} (p) \, \frac{4 \sqrt{2}}{\sqrt{3}} \frac{1}{\beta} Q_2\left(\frac{1}{\beta}\right), \label{eq:matrix_elements_detail2} \\
\langle \mathbf{q} +1, -\mathbf{q} +1 | \hat{M} | n (11) 00 \rangle
&=& - i \left(\frac{4}{3}\right)^2 \frac{\alpha}{2 \pi^{3/2}} \int_0^\infty \frac{\text{d}p \, p^2}{E(p)} \Tilde{R}_{n(11)0} (p) \, 4 \sqrt{2} Q_0\left(\frac{1}{\beta}\right), \label{eq:matrix_elements_detail4} \\
\langle \mathbf{q} +1, -\mathbf{q} +1 | \hat{M} | n (22) 00 \rangle
&=& i \left(\frac{4}{3}\right)^2 \frac{\alpha}{2 \pi^{3/2}} \int_0^\infty \frac{\text{d}p \, p^2}{E(p)} \Tilde{R}_{n(22)0} (p) \, \frac{4 \sqrt{2}}{\sqrt{3}} \frac{1}{\beta} Q_0\left(\frac{1}{\beta}\right), \label{eq:matrix_elements_detail6} 
\end{eqnarray}
while for total photon helicity-2 by
\begin{eqnarray}
\langle \mathbf{q} +1, -\mathbf{q} -1 | \hat{M} | n (02) 22 \rangle
&=& i \left(\frac{4}{3}\right)^2 \frac{\alpha}{2 \pi^{3/2}} \int_0^\infty \frac{\text{d}p \, p^2}{E(p)} \Tilde{R}_{n(02)2} (p) \, \frac{8}{\beta^2} Q_1\left(\frac{1}{\beta}\right), \label{eq:matrix_elements_detail3} \\
\langle \mathbf{q} +1, -\mathbf{q} -1 | \hat{M} | n (20) 22 \rangle
&=& -i \left(\frac{4}{3}\right)^2 \frac{\alpha}{2 \pi^{3/2}} \int_0^\infty \frac{\text{d}p \, p^2}{E(p)} \Tilde{R}_{n(20)2} (p) \, \frac{\sqrt{5}}{\sqrt{2}} \left( \frac{4}{3 \beta^2} -1 \right) \left( \frac{1}{\beta^2} -1 \right) \beta Q^2_2\left(\frac{1}{\beta}\right), \label{eq:matrix_elements_detail5} \\
\langle \mathbf{q} +1, -\mathbf{q} -1 | \hat{M} | n (22) 22 \rangle
&=& -i \left(\frac{4}{3}\right)^2 \frac{\alpha}{2 \pi^{3/2}} \int_0^\infty \frac{ \text{d}p \, p^2}{E(p)} \Tilde{R}_{n(22)2} (p) \, \frac{\sqrt{10}}{\sqrt{7}} \frac{1}{\beta} \left\{ \frac{2}{3} \left( \frac{1}{\beta^2} -1 \right) Q^2_2\left(\frac{1}{\beta}\right) 
- \frac{8}{\beta} Q_1\left(\frac{1}{\beta}\right)  \right\} . \nonumber \\ 
\label{eq:matrix_elements_detail7}
\end{eqnarray}
\end{widetext}
In the above expressions $Q_j(z)$ are the Legendre functions of the second kind, given by 
\begin{equation}
    \begin{aligned}
        Q_0 (z)&= \frac{1}{2} \ln \left( \frac{z+1}{z-1} \right), \\
        Q_1 (z)&= \frac{1}{2} z \ln \left( \frac{z+1}{z-1} \right) -1,\\
        Q_2 (z)&= \frac{1}{4} (3z^2-1) \ln \left( \frac{z+1}{z-1} \right) - \frac{3}{2} z,     \end{aligned}
\end{equation}
for $z>1$, 
and $Q^2_2(z)$ is an associated Legendre functions of second kind, given by: 
\begin{equation}
         Q_2^2 (z)= - \frac{3}{2} (z^2-1) \ln \left( \frac{z+1}{z-1}\right) + \frac{z(3z^2-5)}{z^2 - 1},
\end{equation}
for $z > 1$. 

As the mass of the charmed diquarks is relatively large, it is also instructive to present simplified formulas for the calculation of the two-photon widths of the tetraquark states by making a further non-relativistic approximation in the convolution integrals of Eqs.~\eqref{eq:matrix_elements_detail1}-~\eqref{eq:matrix_elements_detail7}.
This limit requires the momentum to be sufficiently small compared to other scales such as the diquark mass, i.e. $p \ll m_d$ or equivalently $\beta \to 0$. 
Additionally, using the identity
\begin{equation}
     \int \frac{\text{d} p \ p^2}{(2 \pi)^3} p^{L} \Tilde{R}_{n(L S)J} (p)  = \frac{(-i)^L}{4 \pi} \frac{(2L+1)!!}{L!} R_{n(LS)J}^{(L)} (0),\label{eq:originwavefunction}
\end{equation}
where $R_{n(LS)J}^{(L)} (0)$ is the $L$-th derivative of the wavefunction at origin, every integral can be removed from the expression.

We provide a few examples of how the more general expressions of Eqs.~\eqref{eq:matrix_elements_detail1}-~\eqref{eq:matrix_elements_detail7} can be simplified when making such non-relativistic approximation.

Starting with the state $n ^1 S_0$, one uses the limit expansion
\begin{equation}
    \frac{1}{\beta} Q_0(1/\beta) \approx 1 + \mathcal{O}(\beta^2).
\end{equation}
Then, by using Eq.\eqref{eq:originwavefunction} for $L=0$, Eq.~\eqref{eq:matrix_elements_detail1} becomes
\begin{equation}
    \langle \mathbf{q} +1, -\mathbf{q} +1 | \hat{M} | n (00) 00 \rangle = i \left( \frac{4}{3} \right)^2 \frac{\alpha}{m_d} \sqrt{\frac{4 \pi}{3}} R_{n(00)0} (0). \label{eq:s0matrixelement}
\end{equation}
It is straightforward to substitute into Eq.~\eqref{eq:spin0} and get the two-photon width,
\begin{equation}
    \Gamma_{\gamma\gamma} (n^1 S_0) = \left( \frac{4}{3} \right)^4 \frac{\alpha^2}{6 m_d^2} |R_{n(  00)0}(0)|^2.
    \label{eq:s0decaywidth}
\end{equation}

Similarly, Eq.~\eqref{eq:matrix_elements_detail4} for $n ^3 P_0$ can be simplified in this limit by approximating
\begin{equation}
    Q_0 (1/\beta) \approx \beta + \mathcal{O} (\beta^3).
\end{equation}
In this case, the integral in Eq.~\eqref{eq:matrix_elements_detail4} yields the first derivative of the radial wavefunction at the origin using 
Eq.~\eqref{eq:originwavefunction} with $L=1$
\begin{equation}
    \langle \mathbf{q} +1, -\mathbf{q} +1 | \hat{M} | n (11) 00 \rangle = i \left( \frac{4}{3} \right)^2 \frac{12 \alpha}{m_d^2} \sqrt{2 \pi} R^{(1)}_{n(11)0} (0).\label{eq:p0matrixelement}
\end{equation}
Substituting in Eq.~\eqref{eq:spin0} yields
\begin{equation}
    \Gamma_{\gamma\gamma} (n^3 P_0) = \left( \frac{4}{3} \right)^4 \frac{36 \alpha^2}{m_d^4} |R^{(1)}_{n(  11)0}(0)|^2.\label{eq:p0decaywidth}
\end{equation}

The procedure remains similar for the $D$ states as well, with only minor deviations.
For the integral of Eq.~\eqref{eq:matrix_elements_detail5} for the $n ^1 D_2$ states, one can use the non-relativistic expansion  
\begin{equation}
    \left( \frac{4}{3 \beta^2} -1 \right) \left( \frac{1}{\beta^2} -1 \right) \beta Q^2_2\left(\frac{1}{\beta}\right) \approx \frac{32}{15} - \frac{136 \beta^2}{105} + \mathcal{O} (\beta^4). 
\end{equation}
The first term of the expansion does not contribute as it leads to an integral which using Eq.~\eqref{eq:originwavefunction} is proportional to the wavefunction at origin, which vanishes for the $D$-states, $R_{n (2 S) J} (0) =0$.
Consequently, the second term of the expansion is used, which leads to the second derivative of the wavefunction at origin by using Eq.~\eqref{eq:originwavefunction} for $L=2$.
Eq.~\eqref{eq:matrix_elements_detail5} is thus simplified as
\begin{equation}
    \langle \mathbf{q} +1, -\mathbf{q} -1 | \hat{M} | n (20) 22 \rangle = i \left( \frac{4}{3} \right)^2 \frac{34 \alpha}{7 m_d^3} \sqrt{10 \pi} R^{(2)}_{n(20)2} (0).\label{eq:d2matrixelement}
\end{equation}
With the use of Eq.~\eqref{eq:spin2}, one gets
\begin{equation}
    \Gamma_{\gamma\gamma} (n^1 D_2) = \left( \frac{4}{3} \right)^4 \frac{289 \alpha^2}{49 m_d^6} |R^{(2)}_{n(  20)2}(0)|^2.\label{eq:d2decaywidth}
\end{equation}
With the approach presented above, similar simplified equations can be derived for the rest of the states as well.

\bibliography{apssamp}

\providecommand{\noopsort}[1]{}\providecommand{\singleletter}[1]{#1}%
\begin{thebibliography}{28}%
\makeatletter
\providecommand \@ifxundefined [1]{%
 \@ifx{#1\undefined}
}%
\providecommand \@ifnum [1]{%
 \ifnum #1\expandafter \@firstoftwo
 \else \expandafter \@secondoftwo
 \fi
}%
\providecommand \@ifx [1]{%
 \ifx #1\expandafter \@firstoftwo
 \else \expandafter \@secondoftwo
 \fi
}%
\providecommand \natexlab [1]{#1}%
\providecommand \enquote  [1]{``#1''}%
\providecommand \bibnamefont  [1]{#1}%
\providecommand \bibfnamefont [1]{#1}%
\providecommand \citenamefont [1]{#1}%
\providecommand \href@noop [0]{\@secondoftwo}%
\providecommand \href [0]{\begingroup \@sanitize@url \@href}%
\providecommand \@href[1]{\@@startlink{#1}\@@href}%
\providecommand \@@href[1]{\endgroup#1\@@endlink}%
\providecommand \@sanitize@url [0]{\catcode `\\12\catcode `\$12\catcode `\&12\catcode `\#12\catcode `\^12\catcode `\_12\catcode `\%12\relax}%
\providecommand \@@startlink[1]{}%
\providecommand \@@endlink[0]{}%
\providecommand \url  [0]{\begingroup\@sanitize@url \@url }%
\providecommand \@url [1]{\endgroup\@href {#1}{\urlprefix }}%
\providecommand \urlprefix  [0]{URL }%
\providecommand \Eprint [0]{\href }%
\providecommand \doibase [0]{https://doi.org/}%
\providecommand \selectlanguage [0]{\@gobble}%
\providecommand \bibinfo  [0]{\@secondoftwo}%
\providecommand \bibfield  [0]{\@secondoftwo}%
\providecommand \translation [1]{[#1]}%
\providecommand \BibitemOpen [0]{}%
\providecommand \bibitemStop [0]{}%
\providecommand \bibitemNoStop [0]{.\EOS\space}%
\providecommand \EOS [0]{\spacefactor3000\relax}%
\providecommand \BibitemShut  [1]{\csname bibitem#1\endcsname}%
\let\auto@bib@innerbib\@empty
\bibitem [{\citenamefont {{LHCb Collaboration}}(2020)}]{LHCb_2020}%
  \BibitemOpen
  \bibfield  {author} {\bibinfo {author} {\bibnamefont {{LHCb Collaboration}}},\ }\href {https://doi.org/10.1016/j.scib.2020.08.032} {\bibfield  {journal} {\bibinfo  {journal} {Science Bulletin}\ }\textbf {\bibinfo {volume} {65}},\ \bibinfo {pages} {1983–1993} (\bibinfo {year} {2020})}\BibitemShut {NoStop}%
\bibitem [{\citenamefont {Aad}\ \emph {et~al.}(2023)\citenamefont {Aad} \emph {et~al.}}]{ATLAS_2023}%
  \BibitemOpen
  \bibfield  {author} {\bibinfo {author} {\bibfnamefont {G.}~\bibnamefont {Aad}} \emph {et~al.} (\bibinfo {collaboration} {ATLAS}),\ }\bibfield  {journal} {\bibinfo  {journal} {Physical Review Letters}\ }\textbf {\bibinfo {volume} {131}},\ \href {https://doi.org/10.1103/physrevlett.131.151902} {10.1103/physrevlett.131.151902} (\bibinfo {year} {2023})\BibitemShut {NoStop}%
\bibitem [{\citenamefont {Hayrapetyan}\ \emph {et~al.}(2024)\citenamefont {Hayrapetyan} \emph {et~al.}}]{CMS_2024}%
  \BibitemOpen
  \bibfield  {author} {\bibinfo {author} {\bibfnamefont {A.}~\bibnamefont {Hayrapetyan}} \emph {et~al.} (\bibinfo {collaboration} {CMS Collaboration}),\ }\href {https://doi.org/10.1103/PhysRevLett.132.111901} {\bibfield  {journal} {\bibinfo  {journal} {Phys. Rev. Lett.}\ }\textbf {\bibinfo {volume} {132}},\ \bibinfo {pages} {111901} (\bibinfo {year} {2024})}\BibitemShut {NoStop}%
\bibitem [{\citenamefont {Aaboud}\ \emph {et~al.}(2017)\citenamefont {Aaboud} \emph {et~al.}}]{ATLAS_2017}%
  \BibitemOpen
  \bibfield  {author} {\bibinfo {author} {\bibfnamefont {M.}~\bibnamefont {Aaboud}} \emph {et~al.} (\bibinfo {collaboration} {ATLAS}),\ }\href {https://doi.org/10.1038/nphys4208} {\bibfield  {journal} {\bibinfo  {journal} {Nature Phys.}\ }\textbf {\bibinfo {volume} {13}},\ \bibinfo {pages} {852} (\bibinfo {year} {2017})}\BibitemShut {NoStop}%
\bibitem [{\citenamefont {Aad}\ \emph {et~al.}(2021)\citenamefont {Aad} \emph {et~al.}}]{ATLAS_2021}%
  \BibitemOpen
  \bibfield  {author} {\bibinfo {author} {\bibfnamefont {G.}~\bibnamefont {Aad}} \emph {et~al.} (\bibinfo {collaboration} {ATLAS}),\ }\href {https://doi.org/10.1007/JHEP03(2021)243} {\bibfield  {journal} {\bibinfo  {journal} {JHEP}\ }\textbf {\bibinfo {volume} {03}},\ \bibinfo {pages} {243}},\ \bibinfo {note} {[Erratum: JHEP 11, 050 (2021)]}\BibitemShut {NoStop}%
\bibitem [{\citenamefont {Iwasaki}(1975)}]{Iwasaki_1975}%
  \BibitemOpen
  \bibfield  {author} {\bibinfo {author} {\bibfnamefont {Y.}~\bibnamefont {Iwasaki}},\ }\href {https://doi.org/10.1143/PTP.54.492} {\bibfield  {journal} {\bibinfo  {journal} {Progress of Theoretical Physics}\ }\textbf {\bibinfo {volume} {54}},\ \bibinfo {pages} {492} (\bibinfo {year} {1975})}\BibitemShut {NoStop}%
\bibitem [{\citenamefont {Iwasaki}(1976)}]{Iwasaki_1976}%
  \BibitemOpen
  \bibfield  {author} {\bibinfo {author} {\bibfnamefont {Y.}~\bibnamefont {Iwasaki}},\ }\href {https://doi.org/10.1103/PhysRevLett.36.1266} {\bibfield  {journal} {\bibinfo  {journal} {Phys. Rev. Lett.}\ }\textbf {\bibinfo {volume} {36}},\ \bibinfo {pages} {1266} (\bibinfo {year} {1976})}\BibitemShut {NoStop}%
\bibitem [{\citenamefont {Iwasaki}(1977)}]{Iwasaki_1977}%
  \BibitemOpen
  \bibfield  {author} {\bibinfo {author} {\bibfnamefont {Y.}~\bibnamefont {Iwasaki}},\ }\href {https://doi.org/10.1103/PhysRevD.16.220} {\bibfield  {journal} {\bibinfo  {journal} {Phys. Rev. D}\ }\textbf {\bibinfo {volume} {16}},\ \bibinfo {pages} {220} (\bibinfo {year} {1977})}\BibitemShut {NoStop}%
\bibitem [{\citenamefont {Eichten}\ \emph {et~al.}(1975{\natexlab{a}})\citenamefont {Eichten} \emph {et~al.}}]{Cornell_1975}%
  \BibitemOpen
  \bibfield  {author} {\bibinfo {author} {\bibfnamefont {E.}~\bibnamefont {Eichten}} \emph {et~al.},\ }\href {https://doi.org/10.1103/PhysRevLett.34.369} {\bibfield  {journal} {\bibinfo  {journal} {Phys. Rev. Lett.}\ }\textbf {\bibinfo {volume} {34}},\ \bibinfo {pages} {369} (\bibinfo {year} {1975}{\natexlab{a}})}\BibitemShut {NoStop}%
\bibitem [{\citenamefont {Euler}(1936)}]{Euler_1936}%
  \BibitemOpen
  \bibfield  {author} {\bibinfo {author} {\bibfnamefont {H.}~\bibnamefont {Euler}},\ }\href {https://doi.org/https://doi.org/10.1002/andp.19364180503} {\bibfield  {journal} {\bibinfo  {journal} {Annalen der Physik}\ }\textbf {\bibinfo {volume} {418}},\ \bibinfo {pages} {398} (\bibinfo {year} {1936})}\BibitemShut {NoStop}%
\bibitem [{\citenamefont {Heisenberg}\ and\ \citenamefont {Euler}(1936)}]{Heisenberg_1936}%
  \BibitemOpen
  \bibfield  {author} {\bibinfo {author} {\bibfnamefont {W.}~\bibnamefont {Heisenberg}}\ and\ \bibinfo {author} {\bibfnamefont {H.}~\bibnamefont {Euler}},\ }\href {https://doi.org/https://doi.org/10.1007/BF01343663} {\bibfield  {journal} {\bibinfo  {journal} {Zeitschrift für Physik}\ }\textbf {\bibinfo {volume} {98}},\ \bibinfo {pages} {714} (\bibinfo {year} {1936})}\BibitemShut {NoStop}%
\bibitem [{\citenamefont {Pascalutsa}\ and\ \citenamefont {Vanderhaeghen}(2010)}]{Pascalutsa:2010sj}%
  \BibitemOpen
  \bibfield  {author} {\bibinfo {author} {\bibfnamefont {V.}~\bibnamefont {Pascalutsa}}\ and\ \bibinfo {author} {\bibfnamefont {M.}~\bibnamefont {Vanderhaeghen}},\ }\href {https://doi.org/10.1103/PhysRevLett.105.201603} {\bibfield  {journal} {\bibinfo  {journal} {Phys. Rev. Lett.}\ }\textbf {\bibinfo {volume} {105}},\ \bibinfo {pages} {201603} (\bibinfo {year} {2010})}\BibitemShut {NoStop}%
\bibitem [{\citenamefont {Pascalutsa}\ \emph {et~al.}(2012)\citenamefont {Pascalutsa}, \citenamefont {Pauk},\ and\ \citenamefont {Vanderhaeghen}}]{Pascalutsa:2012pr}%
  \BibitemOpen
  \bibfield  {author} {\bibinfo {author} {\bibfnamefont {V.}~\bibnamefont {Pascalutsa}}, \bibinfo {author} {\bibfnamefont {V.}~\bibnamefont {Pauk}},\ and\ \bibinfo {author} {\bibfnamefont {M.}~\bibnamefont {Vanderhaeghen}},\ }\href {https://doi.org/10.1103/PhysRevD.85.116001} {\bibfield  {journal} {\bibinfo  {journal} {Phys. Rev. D}\ }\textbf {\bibinfo {volume} {85}},\ \bibinfo {pages} {116001} (\bibinfo {year} {2012})}\BibitemShut {NoStop}%
\bibitem [{\citenamefont {Biloshytskyi}\ \emph {et~al.}(2022)\citenamefont {Biloshytskyi} \emph {et~al.}}]{Biloshytskyi_2022}%
  \BibitemOpen
  \bibfield  {author} {\bibinfo {author} {\bibfnamefont {V.}~\bibnamefont {Biloshytskyi}} \emph {et~al.},\ }\href {https://doi.org/10.1103/PhysRevD.106.L111902} {\bibfield  {journal} {\bibinfo  {journal} {Phys. Rev. D}\ }\textbf {\bibinfo {volume} {106}},\ \bibinfo {pages} {L111902} (\bibinfo {year} {2022})}\BibitemShut {NoStop}%
\bibitem [{\citenamefont {Debastiani}\ and\ \citenamefont {Navarra}(2019)}]{Debastiani_2019}%
  \BibitemOpen
  \bibfield  {author} {\bibinfo {author} {\bibfnamefont {V.~R.}\ \bibnamefont {Debastiani}}\ and\ \bibinfo {author} {\bibfnamefont {F.~S.}\ \bibnamefont {Navarra}},\ }\href {http://dx.doi.org/10.1088/1674-1137/43/1/013105} {\bibfield  {journal} {\bibinfo  {journal} {Chinese Physics C}\ }\textbf {\bibinfo {volume} {43}},\ \bibinfo {pages} {013105} (\bibinfo {year} {2019})}\BibitemShut {NoStop}%
\bibitem [{\citenamefont {Eichten}\ \emph {et~al.}(1975{\natexlab{b}})\citenamefont {Eichten} \emph {et~al.}}]{Cornell}%
  \BibitemOpen
  \bibfield  {author} {\bibinfo {author} {\bibfnamefont {E.}~\bibnamefont {Eichten}} \emph {et~al.},\ }\href {https://doi.org/10.1103/PhysRevLett.34.369} {\bibfield  {journal} {\bibinfo  {journal} {Phys. Rev. Lett.}\ }\textbf {\bibinfo {volume} {34}},\ \bibinfo {pages} {369} (\bibinfo {year} {1975}{\natexlab{b}})}\BibitemShut {NoStop}%
\bibitem [{\citenamefont {Arnoldi}(1951)}]{Arnoldi_1951}%
  \BibitemOpen
  \bibfield  {author} {\bibinfo {author} {\bibfnamefont {W.~E.}\ \bibnamefont {Arnoldi}},\ }\href {https://doi.org/10.1090/qam/42792} {\bibfield  {journal} {\bibinfo  {journal} {Quarterly of Applied Mathematics}\ }\textbf {\bibinfo {volume} {9}},\ \bibinfo {pages} {17} (\bibinfo {year} {1951})}\BibitemShut {NoStop}%
\bibitem [{\citenamefont {Jacob}\ and\ \citenamefont {Wick}(1959)}]{Jacob_1959}%
  \BibitemOpen
  \bibfield  {author} {\bibinfo {author} {\bibfnamefont {M.}~\bibnamefont {Jacob}}\ and\ \bibinfo {author} {\bibfnamefont {G.~C.}\ \bibnamefont {Wick}},\ }\href {https://doi.org/10.1006/aphy.2000.6022} {\bibfield  {journal} {\bibinfo  {journal} {Annals Phys.}\ }\textbf {\bibinfo {volume} {7}},\ \bibinfo {pages} {404} (\bibinfo {year} {1959})}\BibitemShut {NoStop}%
\bibitem [{\citenamefont {Nachtmann}\ \emph {et~al.}(2006)\citenamefont {Nachtmann} \emph {et~al.}}]{Nachtmann_2005}%
  \BibitemOpen
  \bibfield  {author} {\bibinfo {author} {\bibfnamefont {O.}~\bibnamefont {Nachtmann}} \emph {et~al.},\ }\href {https://doi.org/10.1140/epjc/s2005-02450-3} {\bibfield  {journal} {\bibinfo  {journal} {Eur. Phys. J. C}\ }\textbf {\bibinfo {volume} {45}},\ \bibinfo {pages} {679} (\bibinfo {year} {2006})}\BibitemShut {NoStop}%
\bibitem [{\citenamefont {Danilkin}\ and\ \citenamefont {Vanderhaeghen}(2017{\natexlab{a}})}]{Danilkin_2017}%
  \BibitemOpen
  \bibfield  {author} {\bibinfo {author} {\bibfnamefont {I.}~\bibnamefont {Danilkin}}\ and\ \bibinfo {author} {\bibfnamefont {M.}~\bibnamefont {Vanderhaeghen}},\ }\href {https://doi.org/10.1103/PhysRevD.96.056003} {\bibfield  {journal} {\bibinfo  {journal} {Phys. Rev. D}\ }\textbf {\bibinfo {volume} {96}},\ \bibinfo {pages} {056003} (\bibinfo {year} {2017}{\natexlab{a}})}\BibitemShut {NoStop}%
\bibitem [{\citenamefont {Gerasimov}\ and\ \citenamefont {Moulin}(1975)}]{Gerasimov_1975}%
  \BibitemOpen
  \bibfield  {author} {\bibinfo {author} {\bibfnamefont {S.}~\bibnamefont {Gerasimov}}\ and\ \bibinfo {author} {\bibfnamefont {J.}~\bibnamefont {Moulin}},\ }\href {https://doi.org/https://doi.org/10.1016/0550-3213(75)90438-1} {\bibfield  {journal} {\bibinfo  {journal} {Nuclear Physics B}\ }\textbf {\bibinfo {volume} {98}},\ \bibinfo {pages} {349} (\bibinfo {year} {1975})}\BibitemShut {NoStop}%
\bibitem [{\citenamefont {Brodsky}\ and\ \citenamefont {Schmidt}(1995)}]{Brodsky:1995fj}%
  \BibitemOpen
  \bibfield  {author} {\bibinfo {author} {\bibfnamefont {S.~J.}\ \bibnamefont {Brodsky}}\ and\ \bibinfo {author} {\bibfnamefont {I.}~\bibnamefont {Schmidt}},\ }\bibfield  {title} {\bibinfo {title} {{}},\ }\href {https://doi.org/10.1016/0370-2693(95)00372-R} {\bibfield  {journal} {\bibinfo  {journal} {Phys. Lett. B}\ }\textbf {\bibinfo {volume} {351}},\ \bibinfo {pages} {344} (\bibinfo {year} {1995})}\BibitemShut {NoStop}%
\bibitem [{\citenamefont {Gerasimov}(1965)}]{Gerasimov_1965}%
  \BibitemOpen
  \bibfield  {author} {\bibinfo {author} {\bibfnamefont {S.~B.}\ \bibnamefont {Gerasimov}},\ }\href@noop {} {\bibfield  {journal} {\bibinfo  {journal} {Yad. Fiz.}\ }\textbf {\bibinfo {volume} {2}},\ \bibinfo {pages} {598} (\bibinfo {year} {1965})}\BibitemShut {NoStop}%
\bibitem [{\citenamefont {Drell}\ and\ \citenamefont {Hearn}(1966)}]{Drell_1966}%
  \BibitemOpen
  \bibfield  {author} {\bibinfo {author} {\bibfnamefont {S.~D.}\ \bibnamefont {Drell}}\ and\ \bibinfo {author} {\bibfnamefont {A.~C.}\ \bibnamefont {Hearn}},\ }\href {https://doi.org/10.1103/PhysRevLett.16.908} {\bibfield  {journal} {\bibinfo  {journal} {Phys. Rev. Lett.}\ }\textbf {\bibinfo {volume} {16}},\ \bibinfo {pages} {908} (\bibinfo {year} {1966})}\BibitemShut {NoStop}%
\bibitem [{\citenamefont {Pauk}\ \emph {et~al.}(2013)\citenamefont {Pauk}, \citenamefont {Pascalutsa},\ and\ \citenamefont {Vanderhaeghen}}]{Pauk:2013hxa}%
  \BibitemOpen
  \bibfield  {author} {\bibinfo {author} {\bibfnamefont {V.}~\bibnamefont {Pauk}}, \bibinfo {author} {\bibfnamefont {V.}~\bibnamefont {Pascalutsa}},\ and\ \bibinfo {author} {\bibfnamefont {M.}~\bibnamefont {Vanderhaeghen}},\ }\href {https://doi.org/10.1016/j.physletb.2013.07.058} {\bibfield  {journal} {\bibinfo  {journal} {Phys. Lett. B}\ }\textbf {\bibinfo {volume} {725}},\ \bibinfo {pages} {504} (\bibinfo {year} {2013})}\BibitemShut {NoStop}%
\bibitem [{\citenamefont {Danilkin}\ and\ \citenamefont {Vanderhaeghen}(2017{\natexlab{b}})}]{Danilkin:2016hnh}%
  \BibitemOpen
  \bibfield  {author} {\bibinfo {author} {\bibfnamefont {I.}~\bibnamefont {Danilkin}}\ and\ \bibinfo {author} {\bibfnamefont {M.}~\bibnamefont {Vanderhaeghen}},\ }\href {https://doi.org/10.1103/PhysRevD.95.014019} {\bibfield  {journal} {\bibinfo  {journal} {Phys. Rev. D}\ }\textbf {\bibinfo {volume} {95}},\ \bibinfo {pages} {014019} (\bibinfo {year} {2017}{\natexlab{b}})}\BibitemShut {NoStop}%
\bibitem [{\citenamefont {Ananyev}\ \emph {et~al.}(2020)\citenamefont {Ananyev}, \citenamefont {Danilkin},\ and\ \citenamefont {Vanderhaeghen}}]{Ananyev:2020uve}%
  \BibitemOpen
  \bibfield  {author} {\bibinfo {author} {\bibfnamefont {V.}~\bibnamefont {Ananyev}}, \bibinfo {author} {\bibfnamefont {I.}~\bibnamefont {Danilkin}},\ and\ \bibinfo {author} {\bibfnamefont {M.}~\bibnamefont {Vanderhaeghen}},\ }\href {https://doi.org/10.1103/PhysRevD.102.096019} {\bibfield  {journal} {\bibinfo  {journal} {Phys. Rev. D}\ }\textbf {\bibinfo {volume} {102}},\ \bibinfo {pages} {096019} (\bibinfo {year} {2020})}\BibitemShut {NoStop}%
\bibitem [{\citenamefont {Zhang}\ \emph {et~al.}(2020)\citenamefont {Zhang}, \citenamefont {Ma},\ and\ \citenamefont {Sang}}]{Zhang:2020hoh}%
  \BibitemOpen
  \bibfield  {author} {\bibinfo {author} {\bibfnamefont {H.-F.}\ \bibnamefont {Zhang}}, \bibinfo {author} {\bibfnamefont {Y.-Q.}\ \bibnamefont {Ma}},\ and\ \bibinfo {author} {\bibfnamefont {W.-L.}\ \bibnamefont {Sang}},\ }\href@noop {} {} (\bibinfo {year} {2020}),\ \Eprint {https://arxiv.org/abs/2009.08376} {arXiv:2009.08376 [hep-ph]} \BibitemShut {NoStop}%
\end{thebibliography}%

\end{document}